\documentclass[aoas,preprint]{imsart}

\usepackage{amsthm,amsmath,natbib,graphicx}
\RequirePackage{natbib}
\usepackage{multirow}
\usepackage{multicol}
\usepackage{diagbox}
\usepackage{blkarray}
\usepackage{makecell}
\usepackage{vwcol}  

\usepackage{tikz}
\usetikzlibrary{bayesnet}

\startlocaldefs
\endlocaldefs

\begin{document}

\begin{frontmatter}

\title{Nested Dirichlet Process for population size estimation from multi-list recapture data}
\runtitle{Nested Dirichlet Process for population size estimation from multi-list recapture data}


\author{\fnms{Shuaimin} \snm{Kang}\ead[label=e1]{kang@math.umass.edu}}
\address{\printead{e1}}
\and 
\author{\fnms{Krista} \snm{Gile}\ead[label=e2]{gile@math.umass.edu}}
\and
\address{\printead{e2}}
\author{\fnms{Megan} \snm{Price}\ead[label=e3]{meganp@hrdag.org}}
\address{\printead{e3}}
\runauthor{}

\begin{abstract}
Heterogeneity of response patterns is important in estimating the size of a closed population from multiple recapture data when capture patterns are different over time and location. In this paper, we extend the non-parametric one layer latent class model for multiple recapture data proposed by Manrique-Vallier (2016) to a nested latent class model with the first layer modeling individual heterogeneity and the second layer modeling location-time differences. Location-time groups with similar recording patterns are in the same top layer latent class and individuals within each top layer class are dependent. The nested latent class model incorporates hierarchical heterogeneity into the modeling to estimate population size from multi-list recapture data. This approach leads to more accurate population size estimation and reduced uncertainty.  We apply the method to estimating casualties from the Syrian conflict.
\end{abstract}




\end{frontmatter}

\section{Introduction}
The estimation of the size of a closed population from multi-list recapture data has been studied in many settings, for example estimation of census undercount \citep{Chao1998}\citep{Darroch1993}, estimation of deaths in armed conflict \citep{Ball2003}\citep{Manrique-Vallier2013}\citep{Manrique2019}, estimation of drug injectors \citep{Overstall2014} and estimation of human trafficking victims \citep{Heijden2016}. In general, each record in the multi-list recapture data has descriptive features, like time, location, gender, age, etc. To reduce uncertainty of list capture probabilities imposed by hierarchical structure such as location difference, it's necessary to account for heterogeneity of response patterns in estimating the population size. One way to account for this heterogeneity is stratification. Stratification by every available location or time category may result in too many strata \citep{Ball2003} \citep{Manrique2019}. Combining categories based on expert opinion is highly subjective. In this paper, we put heterogeneity caused by location and time  into the model by building a non-parametric multi-layer latent class model based on the non-parametric one layer latent class model for multiple recapture data (LCMCR) proposed by \cite{Manrique2016}. In Manrique-Vallier (2016)'s paper, the latent layer models individual heterogeneity and individuals in the same latent class are independently captured by each data source. To reflect the hierarchical structure of the data, we add an additional layer on top of the individual layer to capture the top group (location-time) differences and to allow dependence among individuals in the same top latent layer. \\
Our work is motivated by the problem of estimating the number of casualties in the Syrian conflict.  This conflict spanned many years and many regions of Syria, with times and periods of relative calm and heated conflict.  We consider data from four reporting sources, with different reporting patterns across time, region, and type of death. \\
Many techniques estimate the population size by modeling list dependency. A class of generalized linear models, known as log-linear models \citep{Bishop1975} assume the expected log of cell count is linearly related to a set of list interactions. Averaging over Bayesian decomposable graphical models, which represent graphical models of list dependency, is also a classical method to estimate the population size from multi-list recapture data \citep{Madigan1995}\citep{Madigan1997}. Those methods treat all individuals the same which may not be proper in some cases. For example civilian and military deaths in the Syrian conflict data are captured differently by some data lists. Rasch models and extensions on them \citep{Rasch1993} \citep{Darroch1993} \citep{Agresti1994} \citep{Fienberg1999} incorporate individual heterogeneity into the log-linear model. A more flexible method, mixture models has also been used to capture individual heterogeneity \citep{Daniel2008} \citep{Manrique2016}. One strong assumption in the one layer latent class model is that individuals are independent, which might not be proper for data with hierarchical structure.\\
A popular alternative to the one layer latent class model for solving the individual dependence problem in nested data is multi-level latent class models \citep{Vermunt2003} \citep{Yee2006} \citep{Rodrlguez2008}. However, multi-level latent class models haven't been applied in population size estimation for multi-list recapture data. In over-time and across-location multi-list recapture data, we want the top layer to capture location-times that having similar recording patterns and the bottom layer to capture hidden classes of individuals within the top layer latent class. To realize this goal, both the hierarchical Dirichlet process (HDP) \citep{Yee2006} and the nested Dirichlet process (NDP) \citep{Rodrlguez2008} are great candidates. NDP allows both mixture components and weights to change within different top layer classes, but HDP only differs in weights. Due to the complicated and potentially highly distinct list dependencies among top layer latent classes, recording patterns might differ very much between one class containing locations with intense conflicts and one class containing locations with much less conflict. Therefore, we use the nested Dirichlet Process in this paper. NDP is usually applied in clustering nested data, like data with topic hierarchies \citep{Blei2003}\citep{David2010}\citep{Fox2011}. In this paper, we extend it for multiple recapture data to identify more accurate hidden homogeneous classes among top level groups and among individuals within top level groups to better estimate population size from multi-list recapture data.\\
The article is organized as follows. In Section \ref{Motivation_2}, we talk about the data and problem that motivates us for this paper. Then we review the one layer non-parametric latent class model for multiple recapture data (LCMCR), introduce our proposed approach nested latent class model for multiple recapture data (NLCMCR), and we apply MCMC inference for parameter estimation in Section \ref{Model_2}. In Section \ref{Simulation_2}, we do simulations to compare results from the one layer latent class model (LCMCR) and our nested latent class model (NLCMCR). In Section \ref{Application_2}, we apply the NLCMCR in a sample of Syrian conflict data to estimate population size. In Section \ref{Conclusion_2}, we discuss this paper and make conclusions. 
\section{The Syrian conflict data}\label{Motivation_2}
\begin{figure*}
    \caption{Stacked barplots for proportion of records by 15 capture patterns over time in the sampled Syrian conflict data. This plot only shows barplots from four governorates and it's based on monthly data. The recording pattern corresponds to data sources VDC, SNHR, DCHRS, SCSR in order.}
    \label{fig:figure_1}
    \includegraphics[width=.47\textwidth]{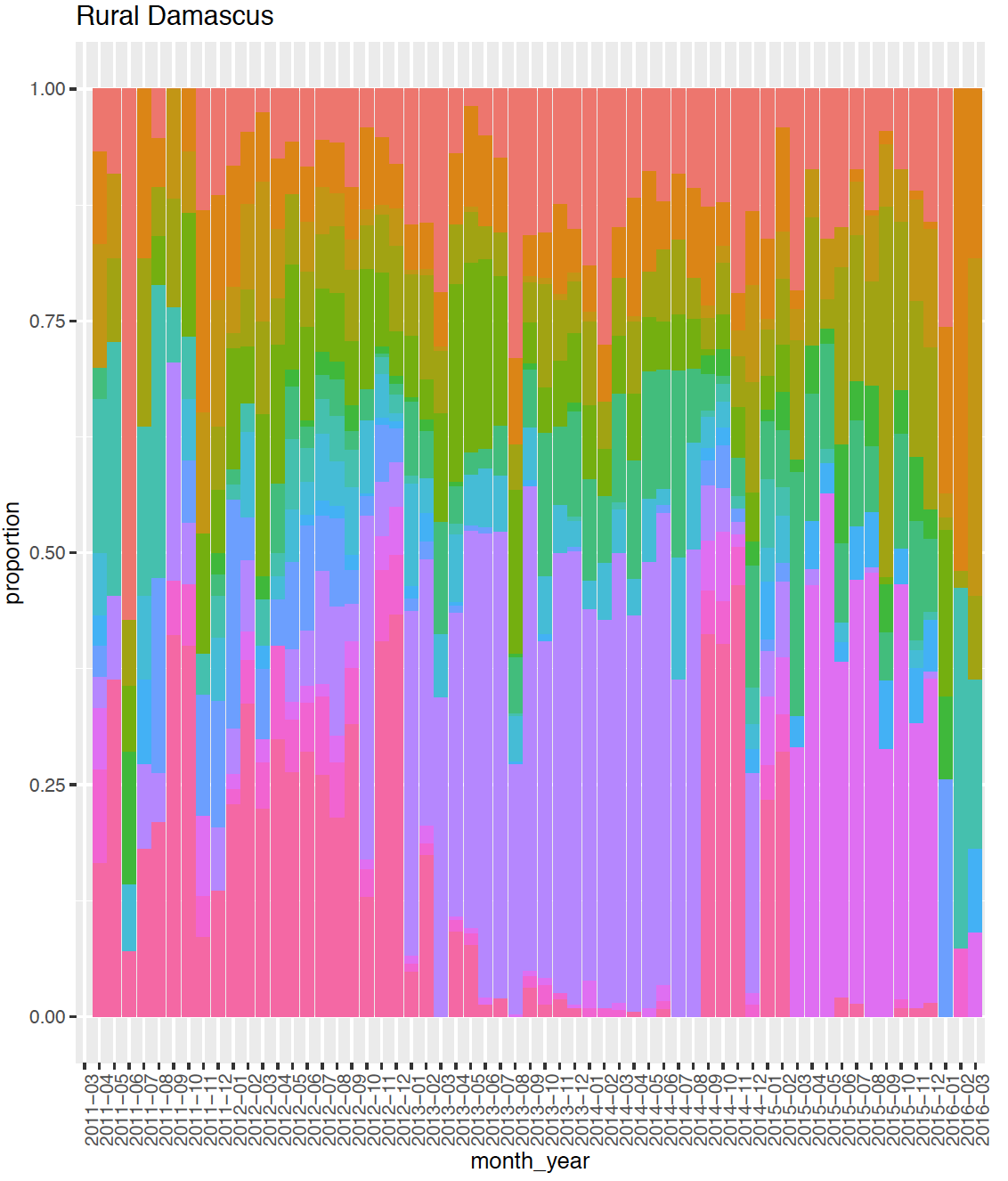}\hfill
    \includegraphics[width=.53\textwidth]{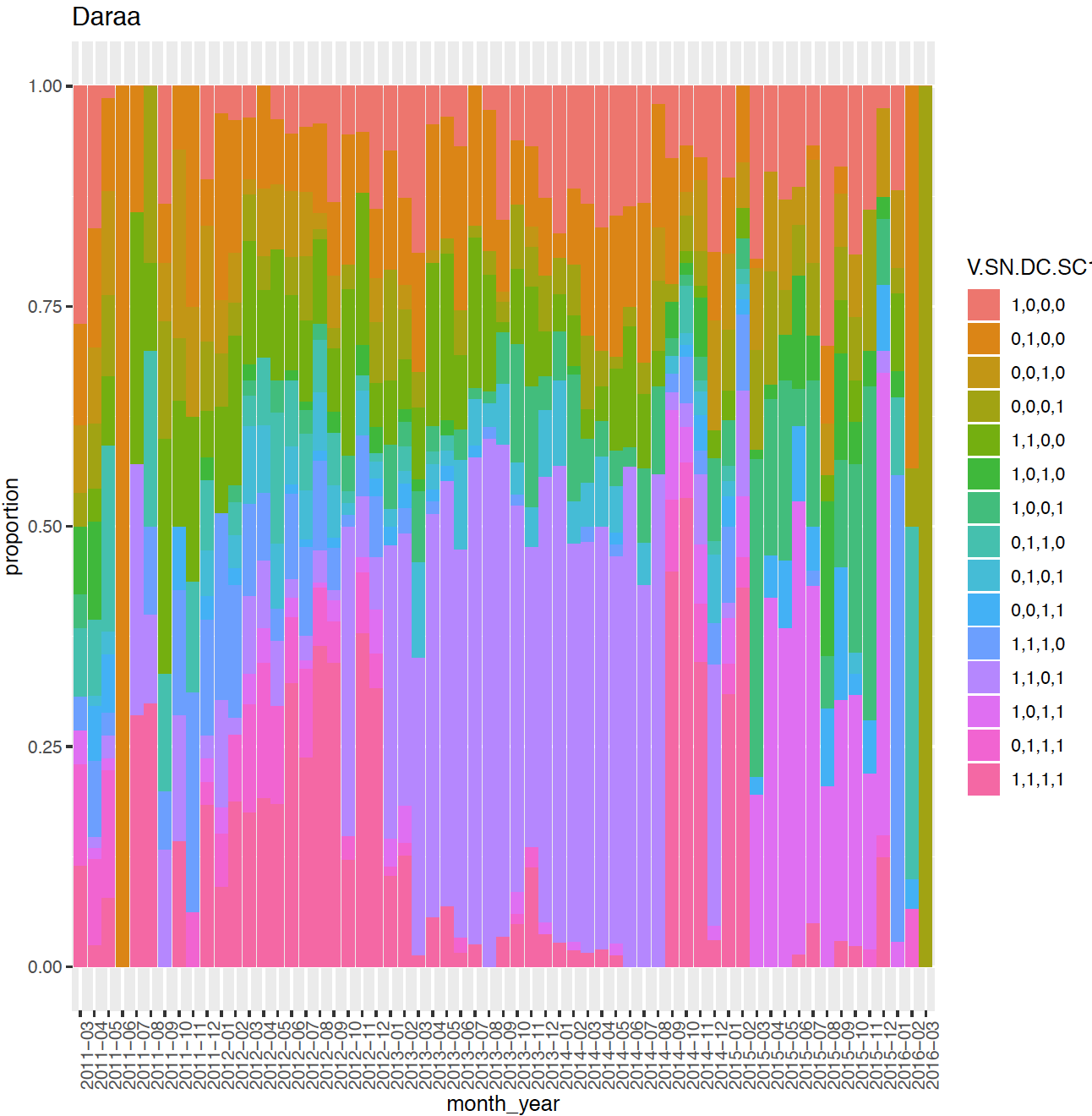}\hfill
    \includegraphics[width=.47\textwidth]{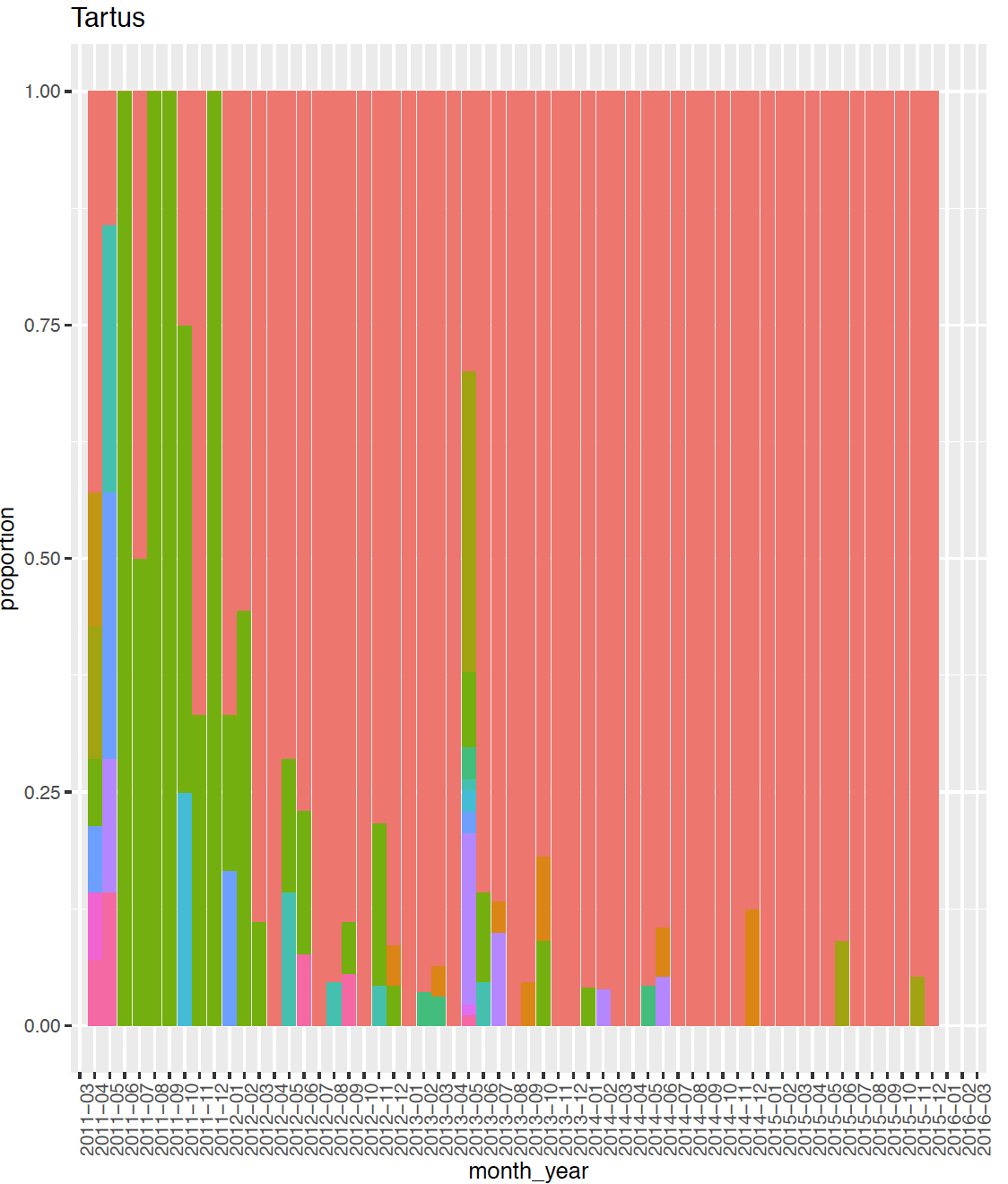}\hfill
     \includegraphics[width=.53\textwidth]{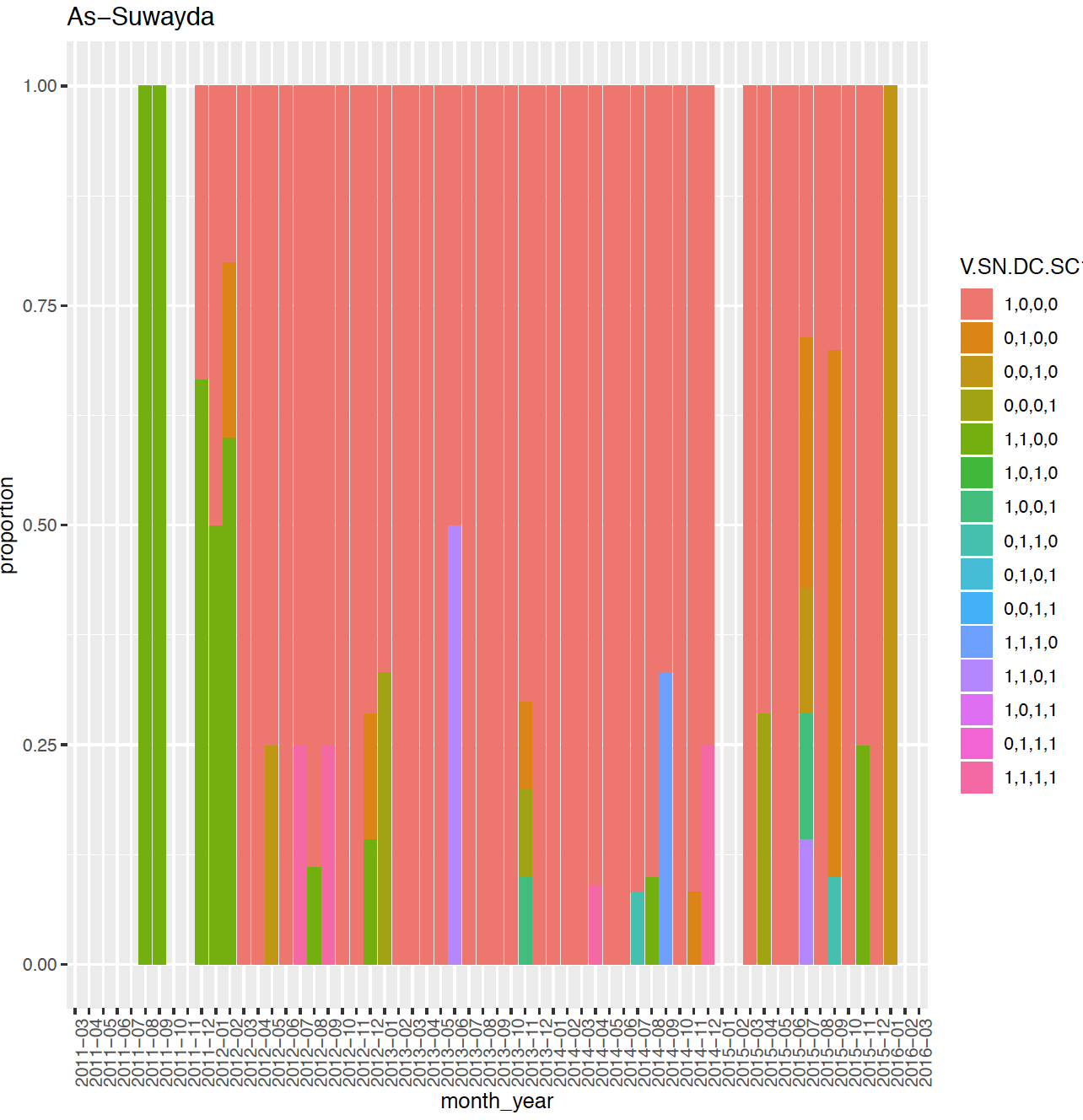}\hfill
\end{figure*}

Human Rights Data Analysis Group (HRDAG) is a non-profit organization that applies rigorous science to the analysis of human rights violations around the world. One of its project is to estimate the total number of killings during the Syrian conflict based on multi-list recapture data. The Syrian conflict data we are using contains documented, identifiable victims who were killed during Syrian conflict from March 2011 to March 2016. Each death record has variables describing this person, which include the person's name, death date, governornate (region in Syria), gender, age. Deaths were recorded by four data sources $(S=4)$ investigating deaths in the Syrian conflict, namely Syrian Center for Statistics and Research (SCSR), Damascus Center for Human Rights Studies (DCHRS), Syrian Network for Human Rights (SNHR) and Violations Documentation Center (VDC). Each record might be captured by more than one data source, thus the number of capture patterns is $2^S - 1 = 15$ excluding the undocumented killings, with $S$ as number of data sources. Due to data confidential, in this paper we randomly generate a sample of $n = 36226$ from all the documented killings. More details about the full documented victims have been discussed in \cite{Price2013}, \cite{price201306}, and \cite{Price2014}. The number of killings recorded under each pattern in this sampled Syrian conflict data is summarized in Table \ref{tab:table_1}. We can see that $n_{1000} = 6039$ deaths are captured by VDC only, $n_{1010} = 652$ are captured by VDC and DCHRS, not by SNHR and SCSR, and $n_{1111} = 4252$ are captured by all four data sources. Estimating the undocumented killings is equivalent to estimating $n_{0000}$, and is the goal of our number inference. 
\begin{table*}
    \centering
    \caption{Number of killings under each recording pattern in the sampled Syrian conflict data}
    \label{tab:table_1}
    \begin{tabular}{|c|c|c|c|c|}
    \hline
           VDC &SNHR &DCHRS &SCSR &Num-Records\\
         \hline
          1 & 0 &0&0 &$n_{1000} = 6039$ \\ 
          0 & 1 &0&0 &$n_{0100} = 3273$ \\
          0 & 0 &1&0 &$n_{0010} = 1363$ \\
          0 & 0 &0&1 &$n_{0001} = 2370$  \\
          1 & 1 &0&0 &$n_{1100} = 3099$  \\
          1 & 0 &1&0 &$n_{1010} = 652$  \\
          1 & 0 &0&1 &$n_{1001} = 2060$  \\
          0 & 1 &1&0 &$n_{0110} = 921$  \\
          0 & 1 &0&1 &$n_{0101} = 1410$  \\
          0 & 0 &1&1 &$n_{0011} = 514$ \\
          1 & 1 &1&0 &$n_{1110} = 1346$ \\
          1 & 1 &0&1 &$n_{1101} = 6483$ \\
          1 & 0 &1&1 &$n_{1011} = 1572$ \\
          0 & 1 &1&1 &$n_{0111} = 872$ \\
          1 & 1 &1&1 &$n_{1111} = 4252$ \\
          \hline
         0 & 0 & 0 & 0 &$n_{0000} = ?$\\
         \hline
    \end{tabular}
\end{table*}
In the Syrian conflict data, documented killings are from 14 governorates across the country. Since our sampled data are generated randomly from the full data set, recording patterns in our sampled data are similar as the full documented Syrian conflict data. From Figure \ref{fig:figure_1}, we can see that recording patterns within governorate change overtime. For example from 04/2011 to 12/2012 deaths captured by all four sources overtake records in other patterns in Rural Damascus. From 01/2013 to 08/2014, more deaths are captured by VDC, SNHR and SCSR together, but not by DCHRS. From 03/2015 to 12/2015, $n_{1101}$ is larger than others or more deaths were captured by VDC, DCHRS and SCSR, but not by SNHR. Some sources capture more killings than others in some governorates, for example, most killings were recorded by VDC in Tartus. Meanwhile, the documented number of killings recorded in different governorates differs much too. All those differences are not hard to explain if we think about the location of each governorate, when and where a small or a big event happened. With those findings, we believe that it's not a good idea to combine all the death records simply over all time and governorates like what we did in Table \ref{tab:table_1} to estimate the total number of killings. Due to the long period and many governorates in this data set, it's also a challenge to do proper stratification subjectively. Therefore, our nested model is important for detecting higher level (e.g. governorate-time) strata in this problem. 
\section{Nested Dirichlet Process for multiple recapture data (NLCMCR)}\label{Model_2}
\subsection{Bayesian non-parametric product-Bernoulli mixture model with Dirichlet Process prior}
\cite{Manrique2016} applied the Bayesian non-parametric latent class model to the population size estimation problem for the multiple recapture data (LCMCR). It assumes the population has some hidden homogeneous strata, within which individuals are captured independently by data lists. The model for it is expressed as the following, 
\begin{align*}
y_{i,s} = \begin{cases}
      1, & \text{if}\ \text{person } i \text{ is captured by the } s^{th} \text{ data list}\\
      0, & \text{otherwise,}
    \end{cases}
\end{align*}
\begin{align*}
    &(y_{i,s}|z_{i} = k) \sim \text{Bernoulli}(\lambda_{k,s})\\
    &z_{i} \sim \text{Cat}(\pi_{1}, \pi_{2},\cdots,\pi_{k},\cdots)\\
    &\lambda_{k,s} \sim \text{Beta}(1,1)\\
    &(\pi_{1},\cdots,\pi_{k},\cdots) \sim \text{SB}(\alpha), \text{ }\alpha \sim \text{Gamma}(a, b)
\end{align*}
where,
\begin{itemize}
    \item $i=1,\cdots,N$, $s=1,\cdots,S$, $k=1,2, \cdots$. $N$ is the population size, $S$ is the number of data lists and $k$ is the latent class label. 
    \item $\lambda_{k,s}$ represents the probability that a person in the $k^{th}$ latent class is captured by the $s^{th}$ data list. 
    \item $z_i$ is the latent class label for the $i^{th}$ individual. It has a categorical prior $Cat(\pi_1, \cdots, \pi_k, \cdots)$.
    \item Prior for the latent class proportion is a stick-breaking prior: $(\pi_1, \cdots, \pi_k, \cdots) \sim SB(\alpha)$. Stick-breaking prior is generally used for non-parametric latent class model to learn number of latent classes instead of specify a certain number to the number of latent classes. 
\end{itemize}
Given prior $P(N) \propto \infty$ to the population size $N$, MCMC can be applied to estimate the population size from re-capture data and list capture probabilities within each homogeneous latent group. \\
This one layer latent class model with Dirichlet Process prior does not account for nested structure in the data. As discussed in \cite{Chen2012}'s paper, ignoring a higher level structure may result in poor classification of individuals to the correct latent class and larger standard errors for estimated group level parameters. Since we can compare latent groups based on individual characteristics, better individual clustering is important in our analysis. Therefore, we extend the one layer latent class model to a multi-level latent class model,  Bayesian non-parametric product-Bernoulli mixture model with nested Dirichlet Process prior. 
\subsection{Bayesian non-parametric product-Bernoulli mixture model with nested Dirichlet Process prior}
\begin{figure*}
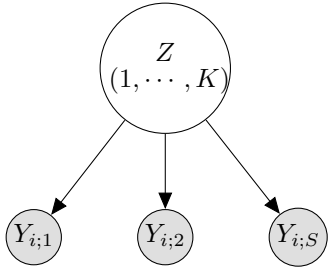

  \centering
  \label{fig:fig_2}
  \caption{One layer Latent class model: $Z$, individual latent class}
  \tikz{ %
    \node[obs] (Y_1) {$Y_{i;1}$} ; %
    \node[obs, right=of Y_1] (Y_2) {$Y_{i;2}$} ; %
    \node[obs, right=of Y_2] (Y_S) {$Y_{i;S}$} ; %
    \node[latent, above=of Y_2,text width=1.5cm,align=center] (Z) {$Z$\\$(1,\cdots,K)$} ; %
    \edge {Z} {Y_1} ; %
    \edge {Z} {Y_2} ; %
    \edge {Z} {Y_S} ; %
  }
  
\end{figure*}

In this paper, we build a nested latent class model, product-Bernoulli mixture model with nested Dirichlet Process as prior, for population size estimation from multi-list recapture data (NLCMCR). Assume individuals belong to latent classes in layer 1, and covariate groups (e.g. location-time) belong to latent classes in layer 2. Conditional on both latent layers, individual capture probabilities for each list are independent, of both other lists and other individuals. The probability that an individual is captured by the $s^{th}$ list is denoted $\lambda_{k,l,s}$, where $k$ is their top-layer class and $l$ is their layer 1 class. This means this probability is influenced by both the individual's layer 1 latent class $l$ and its top level latent class $k$. Meanwhile, for individual $i$ in location-time $j$, its first layer latent class $z^{(1)}_{i,j}$ depends on its top layer latent class $z^{(2)}_j$. From Figure \ref{fig:fig_2}, we can see that individuals in the same latent class are independent given class in the one layer latent class model. From Figure \ref{fig:fig_3}, we can see a graphical model with nested structure. Its top layer latent class reflects group (e.g. location-time) heterogeneity and the first layer models individual heterogeneity within its top layer. In the two layer latent class model, we relax the local independent assumption in the one layer latent class model. Individuals in the same top layer latent class are allowed to be dependent. If our data are given by 
\begin{align*}
y_{i,j,s} = \begin{cases}
      1, & \text{if}\ \text{person } i, \text{in the } j^{th} \text{ top group is captured by the } s^{th} \text{ data list}\\
      0, & \text{otherwise,}
    \end{cases}
\end{align*}
our model is: 
\begin{align*}
    &(y_{i,j,s}|z^{(1)}_{i,j} = l, z^{(2)}_j = k) \sim \text{Bernoulli}(\lambda_{k,l,s})\\
    &(z^{(1)}_{i,j} |z^{(2)}_j = k) \sim \text{Cat}(\pi^{(1)}_{k,1}, \pi^{(1)}_{k,2},\cdots,\pi^{(1)}_{k,l},\cdots)\\
    &z^{(2)}_j \sim \text{Cat}(\pi^{(2)}_{1},\pi^{(2)}_{2}, ,\cdots,\pi^{(2)}_{k}, \cdots)\\
    &\lambda_{k,l,s} \sim \text{Beta}(1,1)\\
    &(\pi^{(1)}_{k,1},\cdots,\pi^{(1)}_{k,l},\cdots) \sim \text{SB}(\alpha_k), \text{ }\alpha_k \sim \text{Gamma}(a_k, b_k)\\
    &(\pi^{(2)}_1,\cdots,\pi^{(2)}_{k},\cdots) \sim \text{SB}(\alpha_0), \text{ }\alpha_0 \sim \text{Gamma}(a_0, b_0),
\end{align*}
 where 
 \begin{itemize}
     \item $i = 1, \cdots, N_j; j = 1, \cdots, J; s = 1, \cdots, S$; $N_j$ and $n_j$ are the number of total and observed individuals in the $j^{th}$ second layer group (e.g. $j^{th}$ location-time), $J$ is the number of second layer groups, S is the number of data sources. Total number of observed individuals is $n = \sum_{j=1}^J n_j$ and the population size is $N = \sum_{j=1}^J N_j$.
     \item $(z^{(1)}_{i,j} = l | z^{(2)}_j = k)$ means the $i^{th}$ person in the $j^{th}$ top group falls into the $l^{th}$ first layer latent class given its second layer latent class as $k$. $k,l = 1, 2, \cdots$.
     \item 
     We use a stick-breaking prior, which is popularly used in non-parametric Bayesian mixture models to learn the number of mixture components from data. 
     $$(\pi^{(2)}_1, \cdots, \pi^{(2)}_k, \cdots) \sim \text{SB}(\alpha_0),\text{ } (\pi^{(1)}_{k,1},\cdots, \pi^{(1)}_{k,l},\cdots,) \sim \text{SB}(\alpha_k),$$
     where 
     $\pi^{(2)}_k = U^{(2)}_k\Pi_{h=1}^{k-1}(1 - U^{(2)}_h)$, $U^{(2)}_k \sim \text{Beta}(1,\alpha_0)$ and $\pi^{(1)}_{k,l} = U^{(1)}_{k,l}\Pi_{h=1}^{l-1}(1 - U^{(1)}_{k,h})$, $U^{(1)}_{k,h} \sim \text{Beta}(1,\alpha_k)$. 
     
      Suppose $(\pi^{(2)}_1, \cdots, \pi^{(2)}_k, \cdots) \sim \text{SB}(\alpha_0)$. For a unit-length stick, each time break a proportion $U^{(2)}$ of the remaining stick. After the $(k-1)^{th}$ break, there is $\Pi_{h=1}^{k-1}(1-U^{(2)}_h)$ left, then the $k^{th}$ break length will be $U^{(2)}_k\Pi_{h=1}^{k-1}(1 - U^{(2)}_h)$, which equals to $\pi^{(2)}_k$. Since $U^{(2)}_k \sim \text{Beta}(1,\alpha_0)$, large $\alpha_0$ gives small break proportions $U^{(2)}_k$ for $k=1, \cdots$, then small break length $\pi^{(2)}_k$ and large number of breaks. Thus, $\alpha_0$ controls the number of latent classes in the second layer and $\alpha_k$ controls the number of latent classes in the first layer given its top layer in latent class $k$. Large $\pi^{(2)}_ks, k=1, \cdots, $ and $\pi^{(1)}_{k,l}s, l = 1, \cdots, $ will corresponding to cluster proportions learnt from data by the model. We take large enough upper bounds $K^*$ and $L^*$ for number of latent classes in the second and first layers.
      \end{itemize}
      
\begin{figure*}
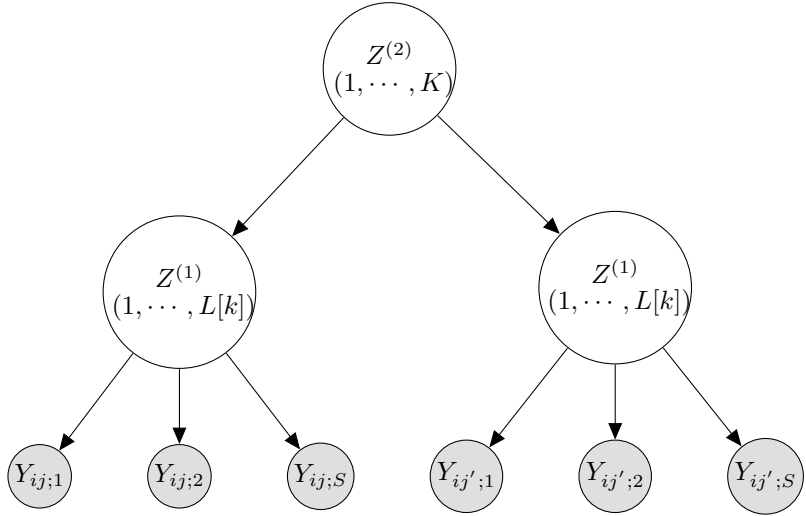

  \centering
  \label{fig:fig_3}
  \caption{Two layers latent class model: $Z^{(1)}$, individual layer, $Z^{(2)}$, top (e.g. location-time) layer.}
  \tikz{ %
    \node[obs] (Y_1j) {$Y_{ij;1}$} ; %
    \node[obs, right=of Y_1j] (Y_2j) {$Y_{ij;2}$} ; %
    \node[obs, right=of Y_2j] (Y_Sj) {$Y_{ij;S}$} ; %
    
    \node[obs, right=of Y_Sj] (Y_1j') {$Y_{ij^{'};1}$} ; 
    \node[obs, right=of Y_1j'] (Y_2j') {$Y_{ij^{'};2}$} ; %
    \node[obs, right=of Y_2j'] (Y_Sj') {$Y_{ij^{'};S}$} ;
    
    \node[latent, above=of Y_2j,text width=1.8cm,align=center] (Z1j) {$Z^{(1)}$\\
    $(1,\cdots,L[k])$} ; %
    \node[latent, above=of Y_2j',text width=1.8cm,align=center] (Z1j') {$Z^{(1)}$\\$(1,\cdots,L[k])$} ; %
    \node[latent, above=of Z1j',xshift=-3cm,text width=1.5cm,align=center] (Z2) {$Z^{(2)}$\\
    $(1,\cdots,K)$} ; %
    
    \edge {Z1j} {Y_1j, Y_2j,Y_Sj} ; %
    \edge {Z1j'} {Y_1j', Y_2j',Y_Sj'} ; %
    \edge {Z2} {Z1j, Z1j'} ; %
  }
  
\end{figure*}

 \subsection{Markov Chain Monte Carlo for parameter estimation}
 An MCMC based Gibbs sampling procedure has been well developed for parameter estimation in the one layer mixture model in the multi-list recapture setting \citep{Manrique2016} \citep{Daniel2008} \citep{Fienberg1999}. Meanwhile, MCMC for the Nested Dirichlet Process is also studied in the clustering nested data problem \citep{Rodrlguez2008}. In this paper, we use the data augmentation and jointly update population size $N$ and latent variables $z^{(2),0}$ and $z^{(1),0}$ using a conditional decomposition \citep{Manrique2016} \citep{Basu2001} to update parameter $N$. For the Nested Dirichlet Process mixture model above, the full likelihood given latent classes $\boldsymbol{z^{(1)},z^{(2)}}$ and parameter set $\Theta = \{\boldsymbol{\lambda,\pi^{(2)},\pi^{(1)}},\alpha_{k=1,\cdots},\alpha_0, a_{k=1,\cdots},b_{k=1,\cdots}, a_0,b_0\}$ is 
 \begin{align*}
     P(\boldsymbol{Y,w}|\boldsymbol{z^{(1)},z^{(2)}},\Theta) &\propto {N \choose{n,w_{1,1},\cdots,w_{k,l},\cdots}}\Pi_{k}\Pi_{l}\big[\pi^{(2)}_{k}\pi^{(1)}_{k,l}\Pi_{s=1}^S (1 - \lambda_{k,l,s})\big]^{w_{k,l}}\\
     &\Pi_{k}\Pi_{l}\Pi_{s=1}^S\big[\pi^{(2)}_{k}\pi^{(1)}_{k,l}(1 - \lambda_{k,l,s})\big]^{n_{k,l,s;0}}\Pi_{k}\Pi_{l}\Pi_{s=1}^S\big[\pi^{(2)}_{k}\pi^{(1)}_{k,l} \lambda_{k,l,s}\big]^{n_{k,l,s;1}}\\
     & I_{(n + \sum_{k}\sum_{l}w_{k,l} = N)},
 \end{align*}
 where $\boldsymbol{w} = \{w_{k,l}; k=1,\cdots,K;l = 1, \cdots,L\}$, $w_{k,l} =$ size of set $\{(y_{i,j,s=1,\cdots,S}=0) \& (z^{(2)}_j=k) \& (z^{(1)}_{i,j}=l)\}$ is the number of un-documented records that fall into the second layer latent class $k$ and the first layer latent class $l$.
     $n_{k,l,s;1} =  ||\{(y_{i,j,s}=1) \& (z^{(2)}_j=k) \& (z^{(1)}_{i,j}=l)\}||$ is the number of documented records falling into the second layer latent class $k$ and the first layer latent class $l$ and captured by the $s^{th}$ data list.
     $n_{k,l,s;0} = ||\{(y_{i,j,s}=0) \& (z^{(2)}_j=k) \& (z^{(1)}_{i,j}=l) \& (y_{i,j,s=1,\cdots,S} \text{ not all equals to 0})\}||$ is the number of documented record fall into the second layer latent class $k$ and the first layer latent class $l$ and not captured by the $s^{th}$ data list.\\
     Instead of setting the number of latent classes to be infinity, truncated approximation is used by setting large numbers $K, L$ to the second and first latent classes \citep{Ishwaran2001}\citep{Ishwaran2002}. The MCMC iterates as follows: 
     
 \begin{enumerate}
 \item Update top layer latent class $z^{(2)}_j$:
 
 \begin{align*}
  &P(z^{(2)}_j=k|Y,\pi^{(1)}) = \sum_{z^{(1)}_{1,j}}\cdots\sum_{z^{(1)}_{n_j,j}}P(z^{(2)}_j=k, z^{(1)}_{1,j},\cdots, z^{(1)}_{n_j,j}|Y)\\
        &\propto \sum_{z^{(1)}_{1,j}}\cdots\sum_{z^{(1)}_{n_j,j}}P(Y|z^{(2)}_j=k, z^{(1)}_{1,j},\cdots, z^{(1)}_{n_j,j})P(z^{(2)}_j=k, z^{(1)}_{1,j},\cdots, z^{(1)}_{n_j,j})\\
        &= \sum_{z^{(1)}_{1,j}}\cdots\sum_{z^{(1)}_{n_j,j}}P(y_{i,j};i=1,\cdots,n_j|z^{(2)}_j=k, z^{(1)}_{1,j},\cdots, z^{(1)}_{n_j,j})\\
        &\hspace{2cm}*P(z^{(1)}_{1,j},\cdots, z^{(1)}_{n_j,j}|z^{(2)}_j=k)P(z^{(2)}_j=k)\\
        &= \sum_{i=1}^{n_j}\sum_{z^{(1)}_{i,j}=1}^{L}\Pi_{s=1}^S\lambda_{k,z^{(1)}_{i,j},s}^{y_{i,j,s}}(1-\lambda_{k,z^{(1)}_{i,j},s})^{1-y_{i,j,s}}\pi^{(1)}_{k,z^{(1)}_{i,j}}\pi^{(2)}_k
    \end{align*}
    
 \item Update first layer latent class
 $z^{(1)}_i$:
 
    \begin{align*}
        &P(z^{(1)}_{i,j}=l|z^{(2)}_j=k, Y,\pi^{(1)}) \\
        &\propto P(y_{ij}|z^{(2)}_j=k,z^{(1)}_{i,j}=l)P(z^{(1)}_i=l|z^{(2)}_j=k)\\
        &\propto \Pi_{s=1}^S\lambda_{k,l,s}^{y_{i,j,s}}(1-\lambda_{k,l,s})^{1-y_{i,j,s}}\pi^{(1)}_{k,l}
    \end{align*}
    
\item Update list capture parameters $\lambda_{k,l,s}$:

    $$P(\lambda_{k,l,s}|\cdots) \propto (1-\lambda_{k,l,s})^{w_{k,l}}\Pi_{j=1}^J\Pi_{i=1}^{n_j}\lambda_{k,l,s}^{y_{i,j,s}}(1-\lambda_{k,l,s})^{1 -y_{i,j,s}}$$
    $(\lambda_{k,l,s}|\cdots)\sim \text{Beta}(1+n_{k,l,s;1}, 1 + n_{k,l,s;0} + w_{k,l}).$
\item Update $\pi^{(2)}_k$: $\pi^{(2)}_k = U^{(2)}_k\Pi_{h<k}(1-U^{(2)}_h)$\\
    Since 
   \begin{align*}
        P(\pi^{(2)}_k, \pi^{(1)}_{k,l}|\cdots) &\propto P(y|\pi^{(2)}_k, \pi^{(1)}_{k,l},\cdots)P(\pi^{(2)}_k, \pi^{(1)}_{k,l})\\ 
        &\propto \Pi_{s=1}^S\big[\pi^{(2)}_k\pi^{(1)}_{k,l}\big]^{n_{k,l,s} + m_{k,l,s} + w_{k,l,s}}\lambda_{k,l,s}^{n_{k,l,s}}(1-\lambda_{k,l,s})^{m_{k,l,s} + w_{k,l,s}}P(\pi^{(2)}_k, \pi^{(1)}_{k,l})
    \end{align*}
    
    changing $\pi^{(2)}_k$ to an expression with $U^{(2)}_k$ using $\pi^{(2)}_k = U^{(2)}_k\Pi_{h<k}(1-U^{(2)}_h)$ and combining with Beta prior of $U^{(2)}_k$, gives a Beta posterior for $U^{(2)}_k$, which we can use the update $\pi^{(2)}_k$.

    let $U^{(2)}_{K^*} = 1$, $U^{(2)}_k \sim \text{Beta}(1 + u^{(2)}_k, \alpha_0 + \sum_{h>k}u^{(2)}_h)$ for $k = 1, \cdots K^{*}-1$, and $u^{(2)}_k = n^{(2)}_k + w^{(2)}_k$. $n^{(2)}_k, w^{(2)}_k$ are the numbers of captured and non-captured individuals whose second layer latent class is $k$. 
    
\item Update $\alpha_0$: $\alpha_0 \sim \text{Gamma}(a_0 - 1 + K^* , b_0  - \text{log}\pi^{(2)}_{K^*})$.
   
\item Update $\pi^{(1)}_{kl}$: $\pi^{(1)}_{kl} = U^{(1)}_{kl}\Pi_{h<l}(1-U^{(1)}_{kh})$\\
        let $U^{(1)}_{kL^*[k]} = 1$, $U^{(1)}_{kl} \sim \text{Beta}(1 + u^{(1)}_{kl}, \alpha_k + \sum_{h>l}u^{(1)}_{kh})$ for $l = 1, \cdots L^{*}[k]-1$, and $u^{(1)}_{kl} = n_{kl} + w_{kl}$. $n_{kl}, w_{kl}$ are the numbers of individuals captured and non-captured in the class with it's first layer latent class $l$ and second layer latent class $k$. 
        
\item Update $\alpha_k$: $\alpha_k \sim \text{Gamma}(a_k - 1 + L^*[k] , b_k  - \text{log}\pi^{(1)}_{kL^*[k]})$.

\item Update $N, w_{kl}$ for all $k,l$: Given $P(N) \propto 1/N$,
        
        $$\small{P(N, w_{kl}|\cdots)\propto \frac{(N-1)!}{\Pi_{k=1}^K\Pi_{l=1}^{L^*[k]}w_{kl}!(n-1)!} \Pi_{k=1}^K\Pi_{l=1}^{L^*[k]}\rho_{kl}^{w_{k1}}(1 - \sum_{k=1}^K\sum_{l=1}^{L^*[k]}\rho_{kl}^{w_{k1}})^{n}}$$
        
        This is a negative multinomial distribution
        with $N = \sum_{k=1}^K\sum_{l=1}^{L^{*}[k]}w_{k1}+n
         = n_0 + n$, $\rho_{kl} = \pi^{(2)}_k\pi^{(1)}_{kl}\Pi_{s=1}^S(1-\lambda_{k,l,s})$.
        Then, 
         $$n_0 \sim \text{NegBinomial}(n, 1 - \sum_{k=1}^K\sum_{l=1}^{L^*[k]}{\pi^{(2)}_k}{\pi^{(1)}_l}\pi_{s=1}^{(2)S}(1-\lambda_{kl;s}))$$
         $$(w_{kl};\textnormal{for all } k,l) \sim \text{Multinomial}(n_0, (p_{kl};\textnormal{for all } k,l))$$
         where $p_{kl} \propto \rho_{kl}$.
        
\end{enumerate}
 
\section{Simulation Study}\label{Simulation_2}
In this section, we generate multiple systems recapture data from a two layer latent class model, then we estimate the population size in three different ways: 
\begin{enumerate}
    \item LCMCR: Latent class model for multiple recapture data.
    \item Multi-LCMCR: Fit LCMCR on each top layer latent class which is known in the simulated data, and then sum up population size estimations for these two sub-groups to get the overall population size.
    \item NLCMCR: nested latent class model for multiple recapture mode.
\end{enumerate}
For the simulated data, we use $S=4$ data sources, $J=100$ (e.g. 100 location-times) top layer groups, $N = 10000$, and number of individuals under each top layer group ($N_j$) ranges from 2 to 602 with a standard deviation of 116. Other parameters for the simulated data are listed in Table~\ref{tab:table_2}. We simulate data by assuming groups within each top layer have similar recording patterns. About $40\%$ of top groups belongs to the first latent class and $60\%$ in the second latent class. From Table~\ref{tab:table_2}, we can see that all four data sources have strong capture probability and they have many overlapping recordings when the top layer latent class is 1 ($z^{(2)} = 1$) and individuals are more likely to be captured by the first data source only or captured by both the first and second data sources when they are in the top layer latent class of 2 ($z^{(2)} = 2$). Therefore, we can see an obvious nested structure in the simulated data. 
\begin{table*}
    \centering
    \caption{Two layer latent class proportions and list capture probabilities in the simulation study}
    \label{tab:table_2}
    \begin{tabular}{|c|c|c|c|c|c|}
\hline
 & &\multicolumn{4}{c|}{List capture probabilities} \\
 \hline
     Layer 2 proportion  &Layer 1 proportion &list 1 & list 2 &list 3 &list 4 \\
     \hline
     \multirow{2}{*}{0.4} & 0.8 &0.9 &0.8 &0.7 &0.6\\
     & 0.2 &0.01 &0.3 &0.1 &0.2\\
     \hline
     \multirow{2}{*}{0.6} & 0.6 &0.1 &0.01 &0.2 &0.05\\
     & 0.4 &0.9 &0.02 &0.1 &0.01\\
     \hline
\end{tabular}
\end{table*}

\begin{figure*}
    \centering
    \caption{Stacked bar-plot of capture pattern proportions by top layer latent class}
    \label{fig:fig_4}
    \includegraphics[scale=0.5]{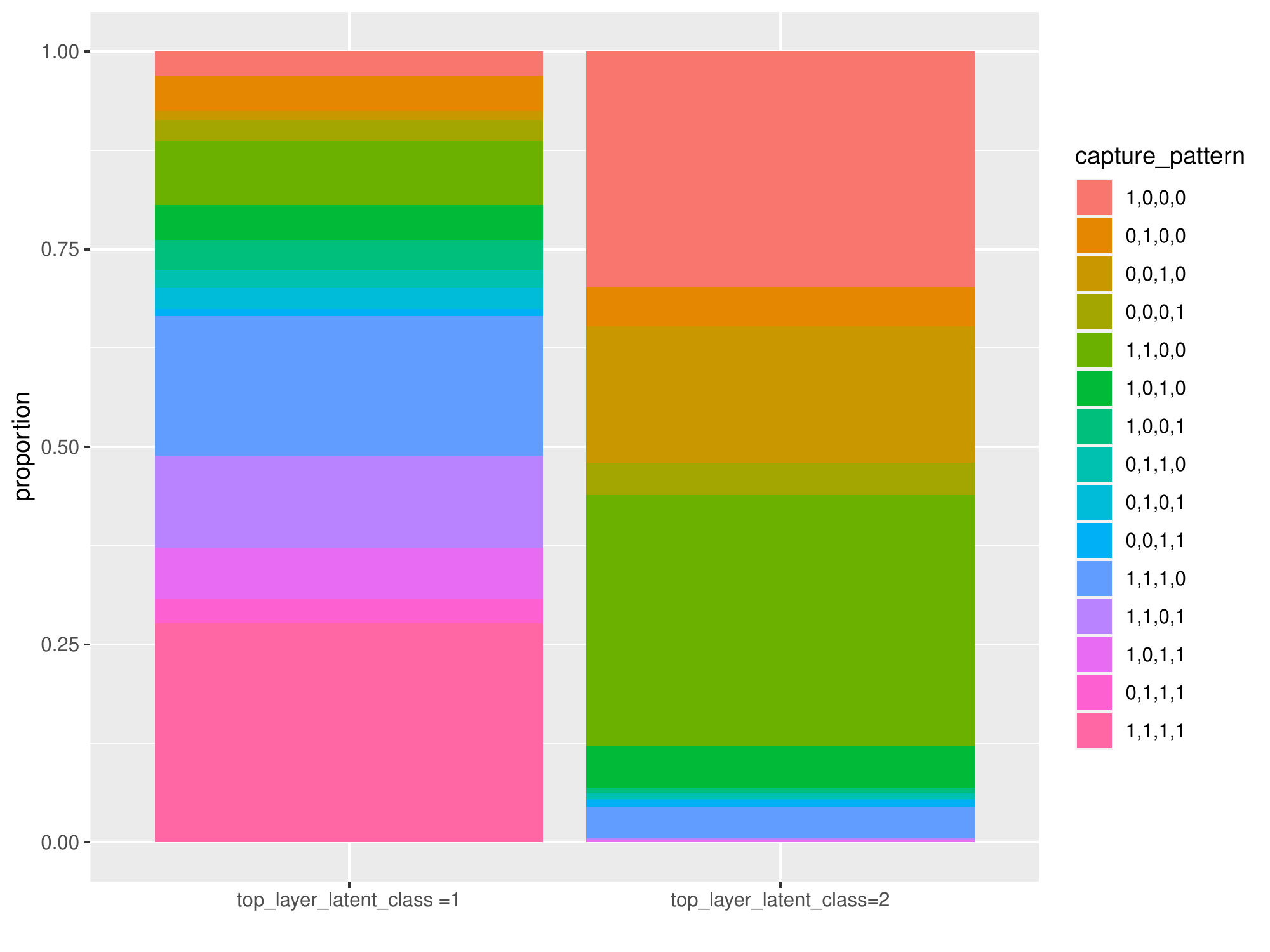}
    
\end{figure*}

\begin{figure*}
    \centering
    \caption{$95\%$ credible intervals for population size estimation under three different models in the simulation study. These boxplots are results from 100 replicates. Boxplots in the middle summarize point estimates, the left and right sets summarize lower and upper boundaries of the $95\%$ credible intervals. The true population size is $N = 10000$ (yellow horizontal line).}
    \label{fig:fig_3}
    \includegraphics[scale=0.6]{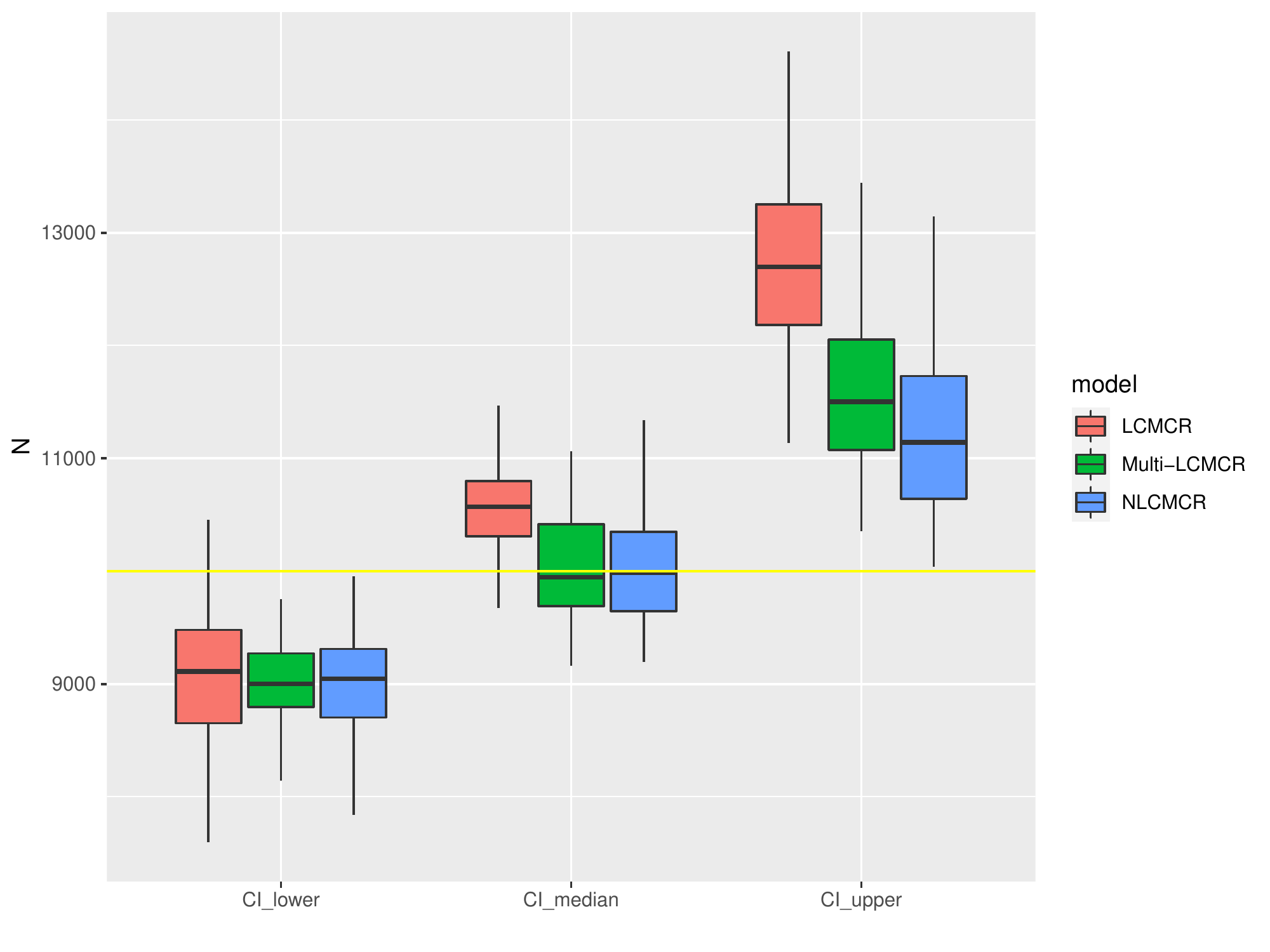}
\end{figure*}

\begin{table}
    \centering
    \caption{\small{Parameter estimates and $95\%$ credible intervals for the simulated data under the nested latent class model (NLMCMCR) and the one-layer latent class model (LCMCR). True values for those parameters are in Table~\ref{tab:table_2}.}}
    \label{tab:simu_params}
    \smallskip\noindent
    \scalebox{0.8}{
    \begin{tabular}{|c|c|c|c|c|c|c|}
\hline
 & & &\multicolumn{4}{c|}{List capture probabilities} \\
 \hline
     Model &Layer 2 proportion  &Layer 1 proportion &list 1 & list 2 &list 3 &list 4\\
     \hline
    \multirow{4}{*}{NLCMCR} &
    \multirow{2}{*}{\makecell{0.38\\(0.26,0.48)}} & \makecell{0.78\\(0.62,0.93)} &\makecell{0.90\\(0.88,0.92)} & \makecell{0.80\\(0.78,0.82)} &\makecell{0.71\\(0.60,0.72)} &\makecell{0.60\\(0.48,0.62)} \\
      & &\makecell{0.21\\(0.05,0.38)} &\makecell{0.18\\(0.11,0.56)} & \makecell{0.34\\(0.22,0.49)} &\makecell{0.12\\(0.06,0.22)} & \makecell{0.22\\(0.05,0.31)} \\\cline{2-7}
    &\multirow{2}{*}{\makecell{0.61\\(0.52,0.73)}} & \makecell{0.62\\(0.54,0.81)} & \makecell{0.11\\(0.06,0.2)} & 
    \makecell{0.01\\(0.01,0.32)} & 
    \makecell{0.19\\(0.11,0.25)} 
    &\makecell{0.05\\(0.03,0.11)}\\ 
    & &\makecell{0.37\\(0.18,0.46)} &\makecell{0.90\\(0.52,0.95)} & \makecell{0.59\\(0.52,0.73)} & \makecell{0.10\\(0.08,0.48)} & \makecell{0.01\\(0.01,0.47)} \\
      \hline
      \multirow{4}{*}{LCMCR} & \multirow{4}{*}{NA} & \makecell{0.53\\(0.39,0.60)} & \makecell{0.80\\(0.50,0.95)} & \makecell{0.96\\(0.69,0.99)} & \makecell{0.84\\(0.66,0.90)} &\makecell{0.94\\(0.81,0.97)}\\
    & &\makecell{0.26\\(0.21,0.27)} &\makecell{0.10\\(0.02,0.36)} & \makecell{0.18\\(0.03,0.57)} & \makecell{0.66\\(0.15,0.94)} & \makecell{0.77\\(0.31,0.98)} \\
    & &\makecell{0.17\\(0.11,0.26)} &\makecell{0.10\\(0.01,0.32)} & \makecell{0.20\\(0.04,0.50)} & \makecell{0.31\\(0.02,0.97)} & \makecell{0.43\\(0.04,0.99)} \\
    & &\makecell{0.03\\(0,0.14)} &\makecell{0.45\\(0.03,0.97)} & \makecell{0.30\\(0.02,0.90)} & \makecell{0.66\\(0.06,0.98)} & \makecell{0.45\\(0.03,0.93)} \\
      \hline
\end{tabular}
}
\end{table}
Figure~\ref{fig:fig_3} is boxplot of posterior estimation of population size under three different models: LCMCR, Multi-LCMCR and NLCMCR. The yellow line is the true population size. We can see that the estimation in LCMCR (red boxplot) is more biased. However, if we use LCMCR to estimate the population size separately by the top layer latent classes, the estimation is much better (green boxplot). This result makes sense because the data within each top layer latent class is from a one layer latent class model. If we stratify the data based on the true top layer latent class, we'll get very good population size estimation using LCMCR for each strata, thus ideal estimation for the overall population size. The estimation from NLCMCR (blue boxplot) is very close to the result from the Multi-LCMCR method. This means that NLCMCR successfully detected the capture pattern differences among top level groups and dependency among individuals in the same top layer latent class. We can also see that NLCMCR gives much smaller uncertainty for the population size estimation than LCMCR.\\
Table \ref{tab:simu_params} summarizes estimates and $95\%$ credible intervals for latent class proportions and list capture probabilities by using the one layer latent class model LCMCR and the nested model NLCMCR. We can see that both LCMCR and NLCMCR learn the number of latent classes as 4 which is the truth. Comparing parameters got from NLCMCR with the true values from which we simulation the data in Table~\ref{tab:table_2}, the estimates are almost unbiased with small uncertainty. To compare parameters got from LCMCR with the true values in Table~\ref{tab:table_2}, we first multiply the top layer proportion by the bottom layer proportion in Table~\ref{tab:table_2} to get the true proportion for the corresponding latent class. We get proportions for the four latent classes $(0.4*0.8 = 0.32, 0.4 * (0.2) = 0.08, 0.6*0.6 = 0.36, 0.6*0.4 = 0.24)$, which differ with the latent class proportions got from LCMCR $(0.53,0.26,0.17,0.03)$ clearly. Since the model LCMCR doesn't group individuals into their true latent classes, the list capture probabilities within each latent class don't reflect the true capture probabilities in Table~\ref{tab:table_2} as well. Overall, both population size estimation and parameter estimation strongly suggest the importance of accounting for hierarchical structure in the multiple recapture data when hierarchy is present in the data. 

\section{Application} \label{Application_2}
\begin{figure*}
    \centering
    \caption{Clustering of the location-time group}
    \label{fig: fig_7}
    \includegraphics[scale=0.6]{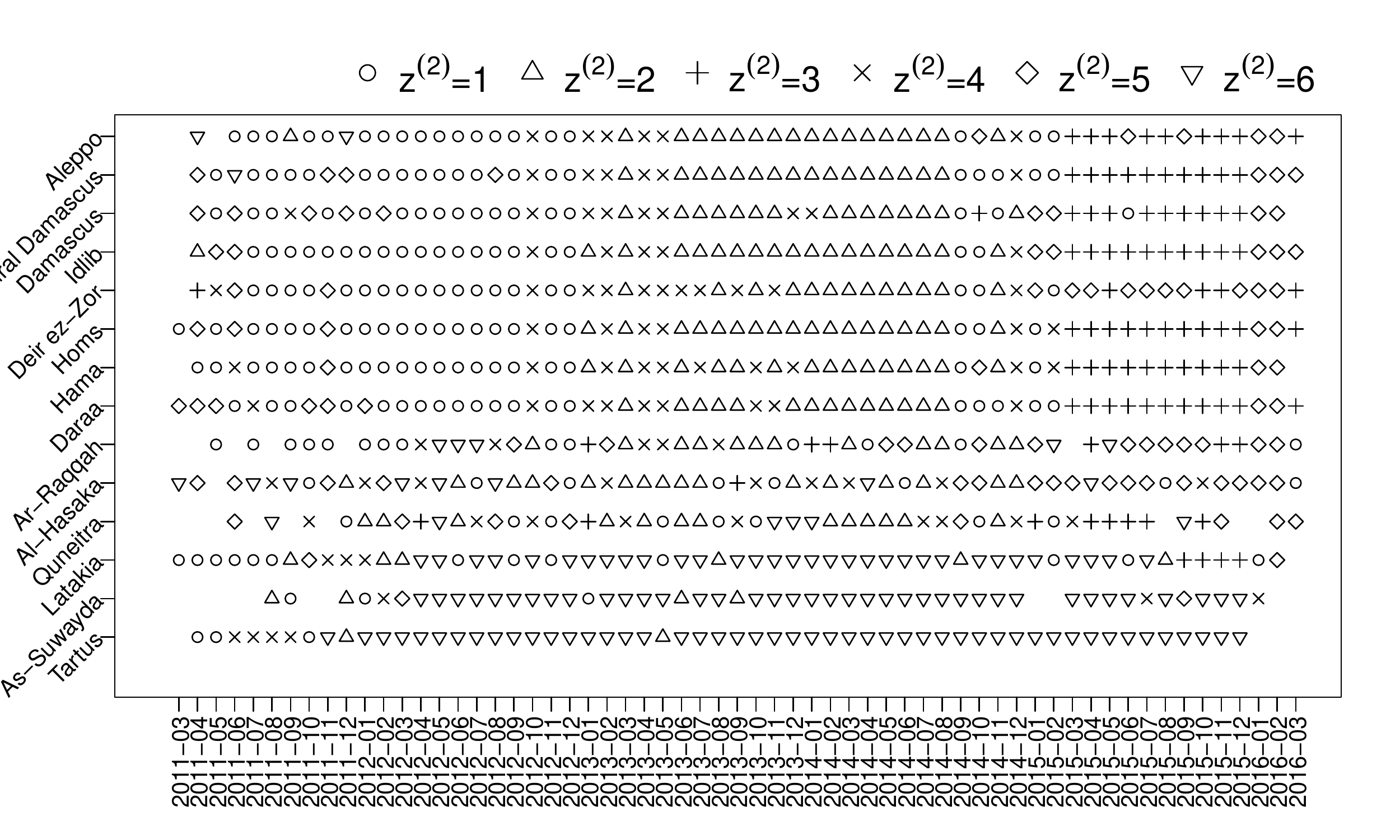}
\end{figure*}

\begin{figure*}
    \centering
    \caption{Proportion of individuals by the individual layer ($z^{(1)}$) for each gov-time within the first gov-time layer ($z^{(2)} = 1, 2$); colored by individual layer, sized by proportion}
    \label{fig: fig_7.1}
    \includegraphics[scale=0.6]{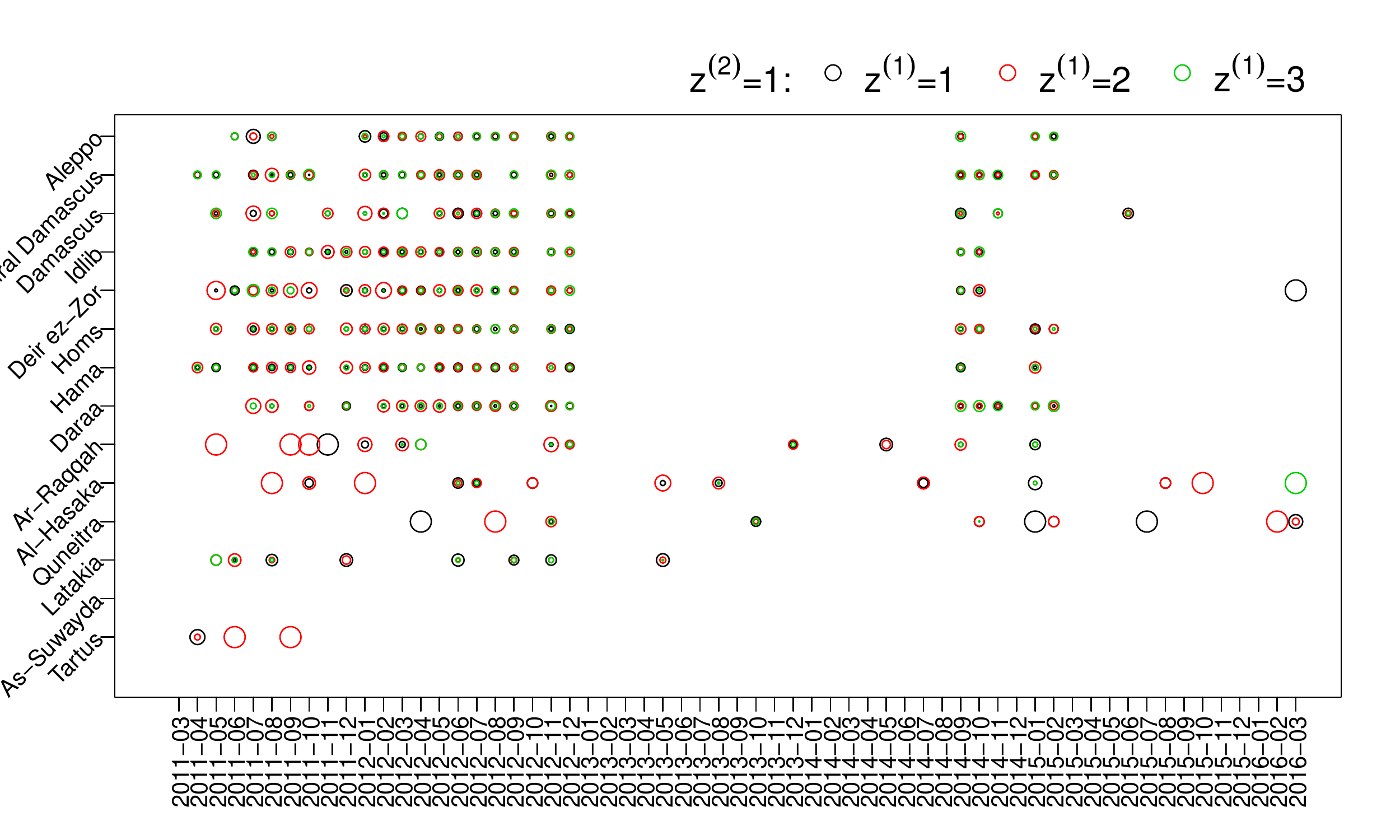}
    \includegraphics[scale=0.6]{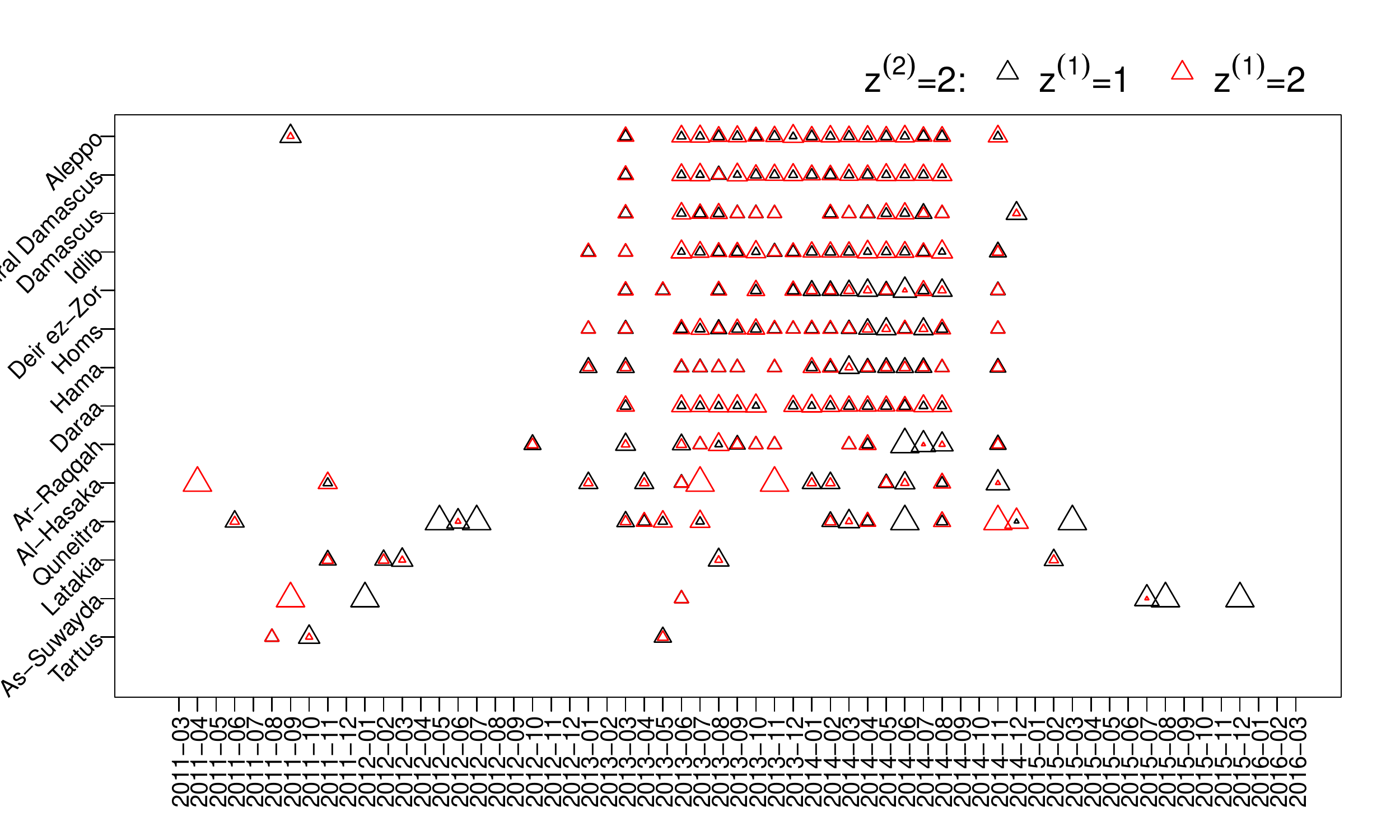}
\end{figure*}

\begin{figure*}
    \centering
    \caption{Proportion of individuals by the individual layer ($z^{(1)}$) for each gov-time within the first gov-time layer ($z^{(2)} = 3, 4$); colored by individual layer, sized by proportion}
    \label{fig: fig_7.2}
    \includegraphics[scale=0.6]{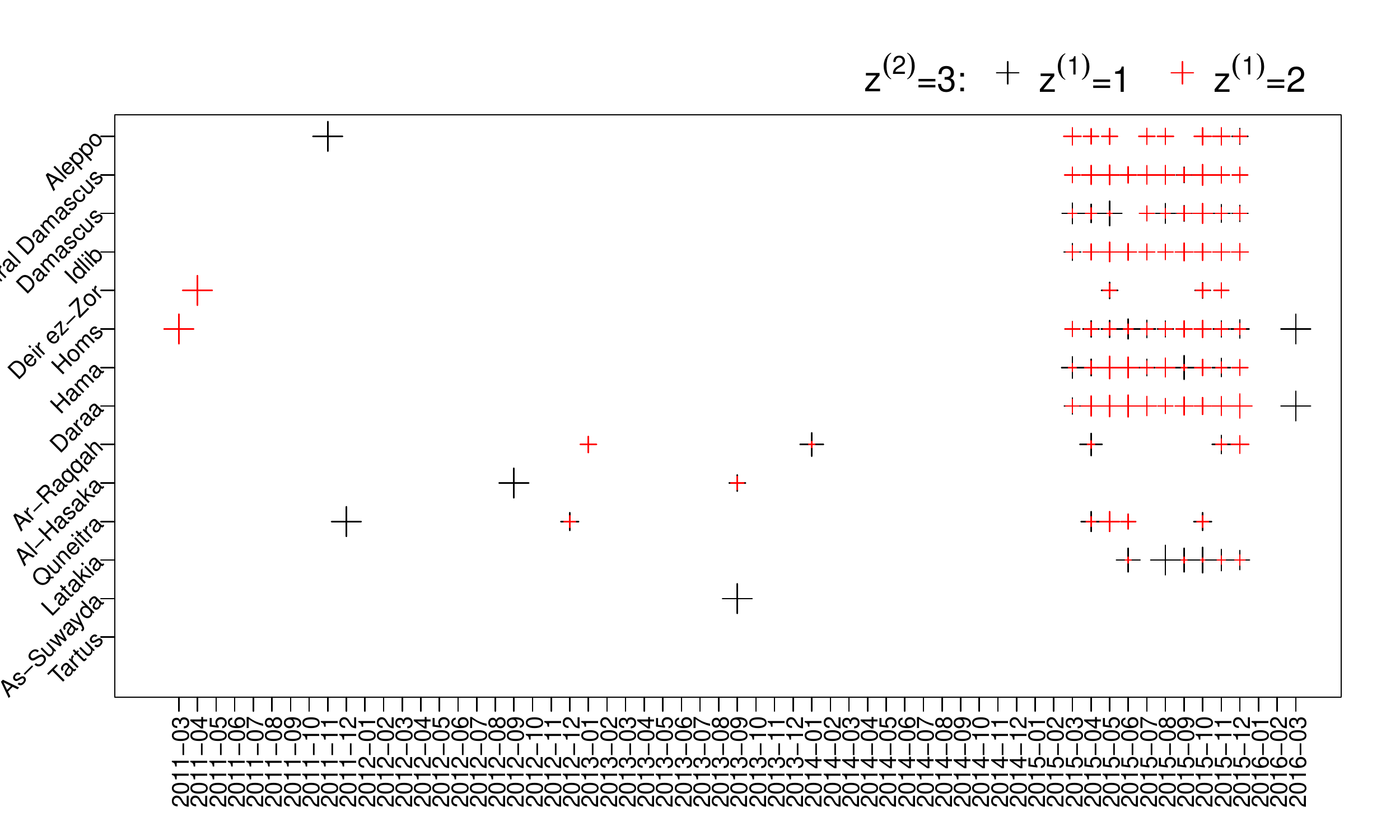}
    \includegraphics[scale=0.6]{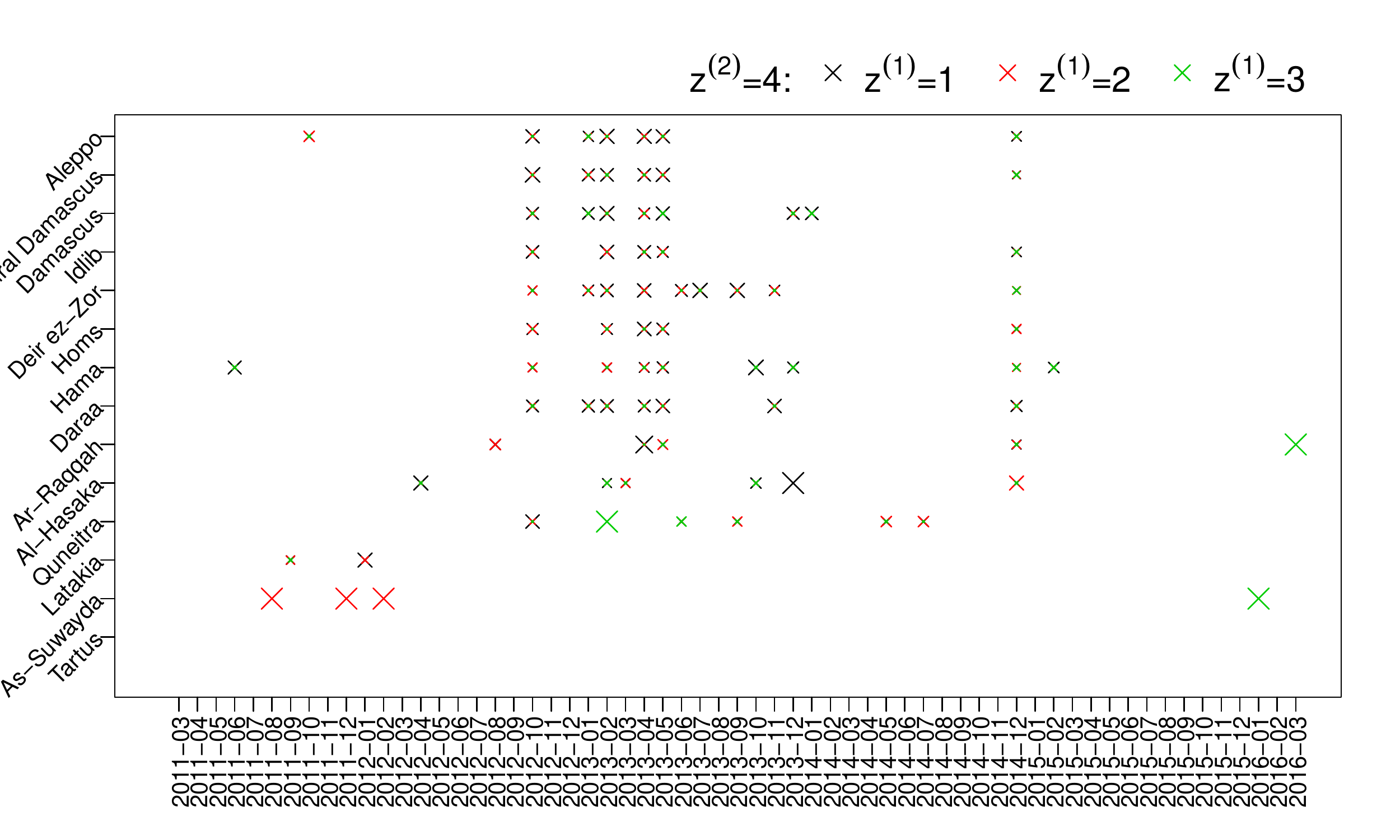}
\end{figure*}

\begin{figure*}
    \centering
    \caption{Proportion of individuals by the individual layer ($z^{(1)}$) for each gov-time within the first gov-time layer ($z^{(2)} = 5, 6$); colored by individual layer, sized by proportion.}
    \label{fig: fig_7.3}
    \includegraphics[scale=0.6]{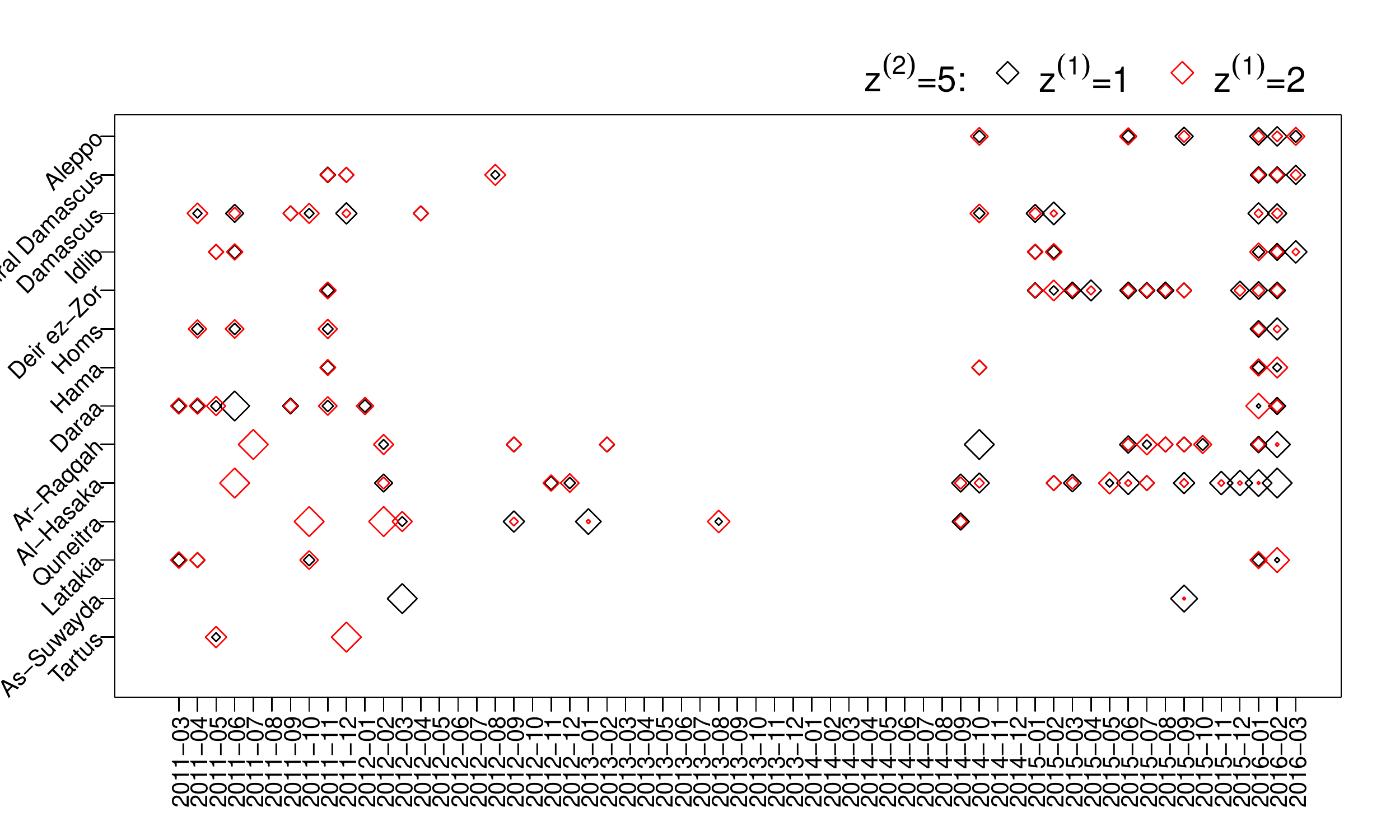}
    \includegraphics[scale=0.6]{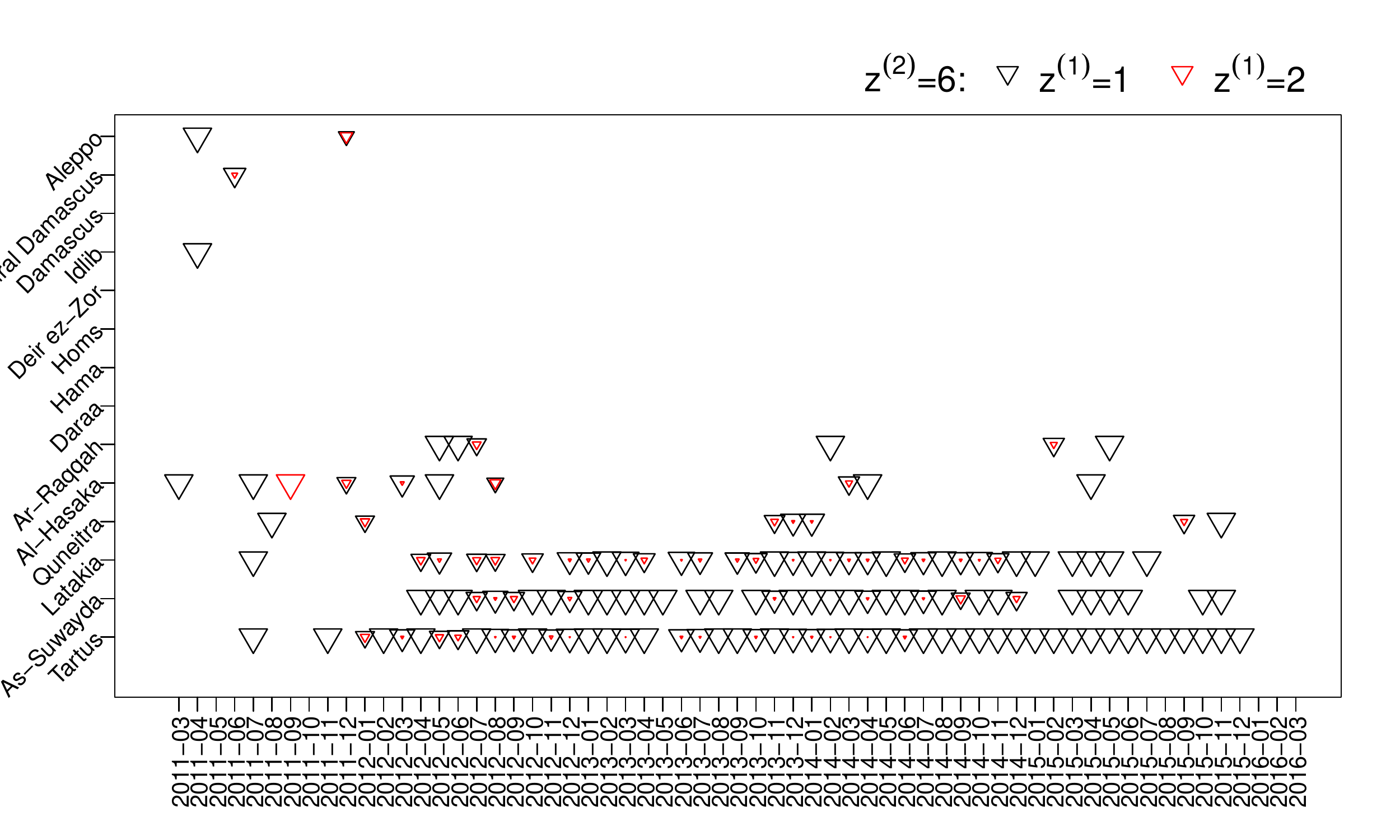}
\end{figure*}

From the descriptions in Section \ref{Motivation_2}, we know that the Syrian conflict multiple systems recapture data has a hierarchical structure over time and governornate. From the simulation results we know that hierarchical structure is important to consider when estimating the population size. Therefore, we take governornate-time as the top group (or second layer) and individuals as the first layer. Then we apply the nested Latent class model for recapture model (NLCMCR) to estimate the total number of killings from the sampled Syrian conflict data. We also apply LCMCR and Bayesian model averaging over decomposable graphical models (BMD)\citep{Madigan1997}\citep{Madigan1995} to compare the results, which are summarized in Table~\ref{tab:table_6}. From Table~\ref{tab:table_6} we see that the population size estimation results from NLCMCR and LCMCR are similar, but NLCMCR has much smaller confidence interval and higher lower credible interval bound. Bayesian model averaging over decomposable graphical model (BMD) gives a much smaller estimation. \\
Figure~\ref{fig: fig_7} summarizes the clustering result of the top layer (governornate-time). We can see that most of the records from Tartus, As-Suwayda and Latakia are clustered into the sixth class ($z^{(2)} = 6$). From Figure~\ref{fig:figure_1} we know that it is because they have similar recording patterns and most deaths recorded in those governorates were from VDC. From Table~\ref{tab:tab_7}, we can also see that individuals in this group are mainly come from the first individual latent class ($z^{(1)} = 1$) with VDC having a much higher capture probability than other data sources (VDC: 0.38, SNHR: 0.038, DCHRS: 0.005, SCSR: 0.011). The second plot in Figure \ref{fig: fig_7.3} gives us some idea about how individuals in each goverornate-time were captured, for example after 06/2014 in Tartus all four data sources have very small capture rate. Governorate-times having red inversed triangles indicates some individuals were captured because of stronger capture probabilities of VDC (0.502) and SNHRS (0.61).\\
We can see that triangles ($z^{(2)} = 2$) in Figure~\ref{fig: fig_7} are mainly from all governorates, excluding Tartus, As-Suwayda and Latakia, from around 03/2013 to 08/2014. Individuals in this group are much less likely to be captured by DCHRS, which is also reflected in Table \ref{tab:tab_7}. Missing individuals in this group are either very likely to be captured by VDC, SNHRS and SCSR together not by DHCRS (VDC: 0.906, SNHRS: 0.841, DCHRS: 0.031, SCSR: 0.882), or very likely to be captured by none of them (VDC: 0.33, SNHRS: 0.254, DCHRS: 0.012, SCSR: 0.16). In the second plot of Figure \ref{fig: fig_7.1}, we can see that red triangles are larger than black triangles in Aleppo, Rural Damasucs, Idlib and Darra which indicates more individual are captured by VDC, SNHRS and SCSR in those governorates.  \\
Similarly, we can summarize that undocumented individuals in Aleppo, Rural Damascus, Damascus, Idlib, Deir ez-Zor, Homs, Hama and Daraa from 03/2011 to 12/2012 (black circled gov-time ($z^{(2)} = 1$) in Figure~\ref{fig: fig_7}) are more likely to be missed by DCHRS (0.457) and SCSR (0.386) or by the four data sources together. In those governorates from 03/2015 to 12/2015 ($z^{(2)} = 3$), most undocumented individuals are either not likely to be captured by all four data sources or only not likely to be captured by SNHRS (0.073). Red plus signs in Aleppo, Rural Damasucs, Idlib and Darra in the first plot of Figure \ref{fig: fig_7.2} dominate, which is similar as the situation in the second plot of Figure \ref{fig: fig_7.1}, but with individuals captured by VDC, DCHRS and SCSR. In those governorates in 10/2012, 01, 02, 04, 05/2013 and 12/2014 ($z^{(2)} = 4$), documented individuals mainly come from individuals who are captured by VDC and SNHRS. Undocumented individuals in gov-times clustered into $z^{(2)} = 5$ are either not likely to be captured by all four data sources or only not likely to be captured by SCSR (0.395).
\begin{table*}
    \centering
    \caption{Estimated number of killings and its $95\%$ posterior credible intervals based on the sampled Syrian Conflict data from 03/2011 to 03/2016}
    \label{tab:table_6}
    \begin{tabular}{|c|c|c|c|c|}
    \hline
   Model & n & $\hat{N}$ & $\hat{N}_{L}$ & $\hat{N}_{U}$ \\
    \hline
         NLCMCR & 36226 & 51447 &48580 &55166 \\
         \hline
         LCMCR &36226 &52070 &46845 &69495 \\
         \hline
         BMD &36226 & 38302 & 36534 &43530 \\
         \hline
    \end{tabular}
\end{table*}

\begin{table*}
    \centering
    \caption{\small{Parameter estimates and $90\%$ credible intervals for the sampled Syrian Conflict data}}
    \label{tab:tab_7}
    \smallskip\noindent
    \scalebox{0.8}{
    \begin{tabular}{|c|c|c|c|c|c|}
\hline
 & &\multicolumn{4}{c|}{List capture probabilities} \\
 \hline
     gov-time layer prop  &individual layer prop &VDC & SNHRS &DCHRS &SCSR\\
     \hline
     \multirow{3}{*}{\makecell{0.333\\(0.3,0.371)}} & \makecell{0.482\\(0.436,0.523)} & \makecell{0.232\\(0.178,0.268)} & \makecell{0.057\\(0.019,0.106)} & \makecell{0.106\\(0.06,0.127)} &\makecell{0.117\\(0.073,0.138)}\\
     &\makecell{0.289\\(0.233,0.319)} &\makecell{0.608\\(0.560,0.889)} & \makecell{0.978\\(0.902,0.998)} & \makecell{0.457\\(0.383,0.899)} & \makecell{0.386\\(0.330,0.884)} \\
     &\makecell{0.219\\(0.161,0.252)} & \makecell{0.894\\(0.612,0.930)} & \makecell{0.914\\(0.897,0.965)} & \makecell{0.98\\(0.463,0.998)} &\makecell{0.952\\(0.352,0.994)} \\
     \hline
     \multirow{2}{*}{\makecell{0.274\\(0.250,0.302)}} & \makecell{0.568\\(0.549,0.599)} &\makecell{0.33\\(0.240,0.424)} & \makecell{0.254\\(0.171,0.335)} &\makecell{0.012\\(0.008,0.018)} &\makecell{0.16\\(0.11,0.222)} \\
      &\makecell{0.431\\(0.390,0.448)} &\makecell{0.906\\(0.886,0.931)} & \makecell{0.841\\(0.821,0.876)} &\makecell{0.031\\(0.027,0.038)} & \makecell{0.882\\(0.839,0.944)} \\
      \hline
      \multirow{2}{*}{\makecell{0.134\\(0.121,0.153)}} & \makecell{0.679\\(0.65,0.713)} &\makecell{0.177\\(0.130,0.243)} & \makecell{0.05\\(0.036,0.079)} &\makecell{0.107\\(0.072,0.150)} &\makecell{0.164\\(0.116,0.223)} \\
      &\makecell{0.321\\(0.284,0.344)} &\makecell{0.867\\(0.716,0.893)} & \makecell{0.073\\(0.062,0.215)} &\makecell{0.733\\(0.260,0.765)} & \makecell{0.815\\(0.430,0.844)} \\
      \hline
      \multirow{3}{*}{\makecell{0.117\\(0.105,0.127)}} & \makecell{0.475\\(0.437,0.687)} &\makecell{0.898\\(0.218,0.919)} & \makecell{0.898\\(0.049,0.919)} &\makecell{0.249\\(0.090,0.270)} &\makecell{0.894\\(0.151,0.929)} \\
      &\makecell{0.321\\(0.284,0.344)} &\makecell{0.144\\(0.038,0.870)} & \makecell{0.263\\(0.078,0.590)} &\makecell{0.110\\(0.011,0.735)} & \makecell{0.225\\(0.040,0.814)} \\
      &\makecell{0.223\\(0,0.263)} &\makecell{0.677\\(0.030,0.903)} & \makecell{0.508\\(0.154,0.626)} &\makecell{0.026\\(0.002,0.566)} & \makecell{0.103\\(0.009,0.536)} \\
      \hline
      \multirow{2}{*}{\makecell{0.079\\(0.057,0.107)}} & \makecell{0.7\\(0.488,0.970)} &\makecell{0.198\\(0.098,0.895)} & \makecell{0.209\\(0.014,0.895)} &\makecell{0.057\\(0.003,0.248)} &\makecell{0.053\\(0.007,0.888)} \\
      &\makecell{0.291\\(0.029,0.373)} &\makecell{0.631\\(0.205,0.764)} & \makecell{0.701\\(0.275,0.952)} &\makecell{0.675\\(0.097,0.888)} & \makecell{0.395\\(0.169,0.647)} \\
      \hline
      \multirow{2}{*}{\makecell{0.062\\(0.035,0.087)}} & \makecell{0.8\\(0.628,0.968)} &\makecell{0.380\\(0.121,0.944)} & \makecell{0.038\\(0.018,0.332)} &\makecell{0.005\\(0.0,0.182)} &\makecell{0.011\\(0.001,0.103)} \\
      &\makecell{0.161\\(0.032,0.370)} &\makecell{0.502\\(0.08,0.747)} & \makecell{0.610\\(0.147,0.941)} &\makecell{0.384\\(0.037,0.862)} & \makecell{0.299\\(0.107,0.649)} \\
      \hline
\end{tabular}
}
\end{table*}


\section{Discussion and Conclusions}\label{Conclusion_2}
In this paper, we find similar capture patterns in some governorates and months in the sampled Syrian conflict recapture data, which means that heterogeneity is clear in this dataset. In order to combine heterogeneity with modeling and allow information sharing across and within strata, we extend the one layer latent class model with Dirichlet Process prior to a multi-layer latent class model with Nested Dirichlet Process prior to estimate population size from multi-list capture data (NLCMCR). In clustering problems, NDP is preferred when the data has a hierarchical structure as it allows dependence for objects within the same top layer latent class. In our multi-list recapture setting, NLCMCR retains its flexible property and uses heterogeneity from both top groups (e.g. location-time) and individuals to detect better latent classes in the data. NLCMCR clusters individuals into more accurate latent classes than LCMCR when the data has a nested structure. This is important in application because it's useful to better summarize group properties based on individual characteristics. Meanwhile, the model NLCMCR gives similar population size estimation with the model LCMCR but with smaller uncertainty if the data has a nested structure. This narrows down uncertainty of our estimation of the true population size. It also reduces uncertainty of other parameter estimates, such as list capture probabilities in each latent class and latent class proportions. Overall, NLCMCR provides both impressed population size estimation and description of the capture patterns. 
\pagebreak
\bibliographystyle{plainnat}
\bibliography{citation}

\begin{thebibliography}{28}
\providecommand{\natexlab}[1]{#1}
\providecommand{\url}[1]{\texttt{#1}}
\expandafter\ifx\csname urlstyle\endcsname\relax
  \providecommand{\doi}[1]{doi: #1}\else
  \providecommand{\doi}{doi: \begingroup \urlstyle{rm}\Url}\fi

\bibitem[Agresti(1994)]{Agresti1994}
Alan Agresti.
\newblock Simple capture-recapture models permitting unequal catchability and
  variable sampling effort.
\newblock \emph{Biometrics}, 50\penalty0 (2):\penalty0 494--500, 1994.
\newblock ISSN 0006341X, 15410420.
\newblock URL \url{http://www.jstor.org/stable/2533391}.

\bibitem[Ball et~al.(2003)Ball, Asher, Sulmont, and Manrique]{Ball2003}
Patrick Ball, Jana Asher, David Sulmont, and Daniel Manrique.
\newblock How many {P}eruvians have died?
\newblock 01 2003.

\bibitem[Basu and Ebrahimi(2001)]{Basu2001}
Sanjib Basu and Nader Ebrahimi.
\newblock Bayesian capture-recapture methods for error detection and estimation
  of population size: Heterogeneity and dependence.
\newblock \emph{Biometrika}, 88\penalty0 (1):\penalty0 269--279, 2001.
\newblock ISSN 00063444.
\newblock URL \url{http://www.jstor.org/stable/2673684}.

\bibitem[Bishop et~al.(1975)Bishop, Holland, and Fienberg]{Bishop1975}
Yvonne~M. Bishop, Paul~W. Holland, and Stephen~E. Fienberg.
\newblock Discrete multivariate analysis: Theory and practice.
\newblock 1975.

\bibitem[Blei et~al.(2003)Blei, Jordan, Griffiths, and Tenenbaum]{Blei2003}
David~M. Blei, Michael~I. Jordan, Thomas~L. Griffiths, and Joshua~B. Tenenbaum.
\newblock Hierarchical topic models and the nested {C}hinese restaurant
  process.
\newblock In \emph{Proceedings of the 16th International Conference on Neural
  Information Processing Systems}, NIPS’03, page 17–24, Cambridge, MA, USA,
  2003. MIT Press.

\bibitem[Blei et~al.(2010)Blei, Griffiths, and Jordan]{David2010}
David~M. Blei, Thomas~L. Griffiths, and Michael~I. Jordan.
\newblock The nested {C}hinese restaurant process and {B}ayesian nonparametric
  inference of topic hierarchies.
\newblock \emph{J. ACM}, 57\penalty0 (2), 02 2010.
\newblock ISSN 0004-5411.
\newblock URL \url{https://doi.org/10.1145/1667053.1667056}.

\bibitem[Chao and Tsay(1998)]{Chao1998}
Anne Chao and P.~K. Tsay.
\newblock A sample coverage approach to multiple-system estimation with
  application to census undercount.
\newblock \emph{Journal of the American Statistical Association}, 93\penalty0
  (441):\penalty0 283--293, 1998.

\bibitem[Chen(2012)]{Chen2012}
Qi~Chen.
\newblock The impact of ignoring a level of nesting structure in multilevel
  mixture model: A {M}onte {C}arlo study.
\newblock \emph{SAGE Open}, 2\penalty0 (1):\penalty0 2158244012442518, 2012.

\bibitem[Darroch et~al.(1993)Darroch, Fienberg, Glonek, and
  Junker]{Darroch1993}
John~N. Darroch, Stephen~E. Fienberg, Gary F.~V. Glonek, and Brian~W. Junker.
\newblock A three-sample multiple-recapture approach to census population
  estimation with heterogeneous catchability.
\newblock \emph{Journal of the American Statistical Association}, 88\penalty0
  (423):\penalty0 1137--1148, 1993.

\bibitem[Fienberg et~al.(1999)Fienberg, Johnson, and Junker]{Fienberg1999}
S.~E. Fienberg, M.~S. Johnson, and B.~W. Junker.
\newblock Classical multilevel and {B}ayesian approaches to population size
  estimation using multiple lists.
\newblock \emph{Journal of the Royal Statistical Society: Series A (Statistics
  in Society)}, 162\penalty0 (3):\penalty0 383--405, 1999.

\bibitem[Fox et~al.(2011)Fox, Sudderth, Jordan, and Willsky]{Fox2011}
Emily~B. Fox, Erik~B. Sudderth, Michael~I. Jordan, and Alan~S. Willsky.
\newblock A sticky {HDP-HMM} with application to speaker diarization.
\newblock \emph{The Annals of Applied Statistics}, 5\penalty0 (2A):\penalty0
  1020--1056, 2011.
\newblock ISSN 19326157.
\newblock URL \url{http://www.jstor.org/stable/23024915}.

\bibitem[Heijden(2016)]{Heijden2016}
Peter~G.M. Heijden.
\newblock Multiple systems estimation for estimating the number of victims of
  human trafficking across the world.
\newblock 06 2016.

\bibitem[Ishwaran and James(2001)]{Ishwaran2001}
Hemant Ishwaran and Lancelot~F James.
\newblock Gibbs sampling methods for {S}tick-breaking priors.
\newblock \emph{Journal of the American Statistical Association}, 96\penalty0
  (453):\penalty0 161--173, 2001.

\bibitem[Ishwaran and James(2002)]{Ishwaran2002}
Hemant Ishwaran and Lancelot~F. James.
\newblock Approximate {D}irichlet process computing in finite normal mixtures:
  Smoothing and prior information.
\newblock \emph{Journal of Computational and Graphical Statistics}, 11\penalty0
  (3):\penalty0 508--532, 2002.
\newblock ISSN 10618600.
\newblock URL \url{http://www.jstor.org/stable/1391111}.

\bibitem[Madigan and York(1997)]{Madigan1997}
David Madigan and Jeremy~C. York.
\newblock Bayesian methods for estimation of the size of a closed population.
\newblock \emph{Biometrika}, 84\penalty0 (1):\penalty0 19--31, 1997.
\newblock ISSN 00063444.
\newblock URL \url{http://www.jstor.org/stable/2337552}.

\bibitem[Madigan et~al.(1995)Madigan, York, and Allard]{Madigan1995}
David Madigan, Jeremy York, and Denis Allard.
\newblock Bayesian graphical models for discrete data.
\newblock \emph{International Statistical Review / Revue Internationale de
  Statistique}, 63\penalty0 (2):\penalty0 215--232, 1995.
\newblock ISSN 03067734, 17515823.
\newblock URL \url{http://www.jstor.org/stable/1403615}.

\bibitem[Manrique-Vallier(2016)]{Manrique2016}
Daniel Manrique-Vallier.
\newblock Bayesian population size estimation using {D}irichlet process
  mixtures.
\newblock \emph{Biometrics}, 72\penalty0 (4):\penalty0 1246--1254, 2016.

\bibitem[Manrique-Vallier and Fienberg(2008)]{Daniel2008}
Daniel Manrique-Vallier and Stephen~E. Fienberg.
\newblock Population size estimation using individual level mixture models.
\newblock \emph{Biometrical Journal}, 50\penalty0 (6):\penalty0 1051--1063,
  2008.

\bibitem[Manrique-Vallier et~al.(2013)Manrique-Vallier, Price, and
  Gohdes]{Manrique-Vallier2013}
Daniel Manrique-Vallier, Megan Price, and Anita Gohdes.
\newblock \emph{Multiple Systems Estimation Techniques for Estimating
  Casualties in Armed Conflicts}, pages 165--181.
\newblock 06 2013.

\bibitem[{Manrique-Vallier} et~al.(2019){Manrique-Vallier}, {Ball}, and
  {Sulmont}]{Manrique2019}
Daniel {Manrique-Vallier}, Patrick {Ball}, and David {Sulmont}.
\newblock {Estimating the number of fatal victims of the Peruvian internal
  armed conflict, 1980-2000: an application of modern multi-list
  Capture-Recapture techniques}.
\newblock \emph{arXiv e-prints}, art. arXiv:1906.04763, Jun 2019.

\bibitem[Overstall et~al.(2014)Overstall, King, Bird, Hutchinson, and
  Hay]{Overstall2014}
Antony~M. Overstall, Ruth King, Sheila~M. Bird, Sharon~J. Hutchinson, and
  Gordon Hay.
\newblock Incomplete contingency tables with censored cells with application to
  estimating the number of people who inject drugs in {S}cotland.
\newblock \emph{Statistics in Medicine}, 33\penalty0 (9):\penalty0 1564--1579,
  2014.

\bibitem[Price et~al.(2013{\natexlab{a}})Price, Klingner, and Ball]{Price2013}
Megan Price, Jeff Klingner, and Patrick Ball.
\newblock Preliminary statistical analysis of documentation of killings in the
  {S}yrian {A}rab {R}epublic.
\newblock \emph{The Benetech Human Rights Program, commissioned by the United
  Nations Office of the High Commissioner for Human Rights (OHCHR)}, 01
  2013{\natexlab{a}}.

\bibitem[Price et~al.(2013{\natexlab{b}})Price, Klingner, Qtiesh, and
  Ball]{price201306}
Megan Price, Jeff Klingner, Anas Qtiesh, and Patrick Ball.
\newblock Full updated statistical analysis of documentation of killings in the
  {S}yrian {A}rab {R}epublic.
\newblock \emph{Human Rights Data Analysis Group, commissioned by the United
  Nations Office of the High Commissioner for Human Rights (OHCHR)}, 06
  2013{\natexlab{b}}.

\bibitem[Price et~al.(2014)Price, Gohdes, and Ball]{Price2014}
Megan Price, Anita Gohdes, and Patrick Ball.
\newblock Updated statistical analysis of documentation of killings in the
  {S}yrian {A}rab {R}epublic.
\newblock \emph{Human Rights Data Analysis Group, commissioned by the United
  Nations Office of the High Commissioner for Human Rights (OHCHR)}, 08 2014.

\bibitem[Rasch(1993)]{Rasch1993}
G.E. Rasch.
\newblock \emph{Probabilistic Models for Some Intelligence and Attainment
  Tests}, volume~1.
\newblock 01 1993.

\bibitem[Rodrlguez et~al.(2008)Rodrlguez, Dunson, and Gelfand]{Rodrlguez2008}
Abel Rodrlguez, David~B Dunson, and Alan~E Gelfand.
\newblock The nested {D}irichlet process.
\newblock \emph{Journal of the American Statistical Association}, 103\penalty0
  (483):\penalty0 1131--1154, 2008.

\bibitem[Teh et~al.(2006)Teh, Jordan, Beal, and Blei]{Yee2006}
Yee~Whye Teh, Michael~I Jordan, Matthew~J Beal, and David~M Blei.
\newblock Hierarchical {D}irichlet processes.
\newblock \emph{Journal of the American Statistical Association}, 101\penalty0
  (476):\penalty0 1566--1581, 2006.

\bibitem[Vermunt(2003)]{Vermunt2003}
Jeroen~K. Vermunt.
\newblock Multilevel latent class models.
\newblock \emph{Sociological Methodology}, 33\penalty0 (1):\penalty0 213--239,
  2003.

\end{thebibliography}
\end{document}